\documentclass[a4paper,11pt]{article}
\pdfoutput=1 

\usepackage{jheppub-mod} 

\usepackage[T1]{fontenc} 
\usepackage{xcolor}
\usepackage{dsfont}
\usepackage{slashed}

\newcommand{\cC}{{\cal C}}
\newcommand{\cD}{{\cal D}}

\newcommand{\cF}{{\cal F}}
\newcommand{\cG}{{\cal G}}
\newcommand{\cH}{{\cal H}}

\newcommand{\cL}{{\cal L}}

\newcommand{\cN}{{\cal N}}

\newcommand{\be}{\begin{equation}}
\newcommand{\ee}{\end{equation}}
\newcommand{\bea}{\begin{eqnarray}}
\newcommand{\eea}{\end{eqnarray}}

\newcommand{\pr}{\partial}

\newcommand{\bPsi}{\bar{\Psi}}

\title{Carroll fermions of arbitrary spin}

\author[a,1]{Andrea Campoleoni\note{Research Associate of the Fund for Scientific Research -- FNRS, Belgium.}}
\author[a,b,2]{and Lea Mele\note{FRIA grantee of the Fund for Scientific Research -- FNRS, Belgium.}}

\affiliation[a]{Service de Physique de l’Univers, Champs et Gravitation,\\
Universit\'e de Mons -- UMONS, 20 place du Parc, 7000 Mons, Belgium}

\affiliation[b]{Erwin Schr\"odinger International Institute for Mathematics and Physics,\\ University of Vienna, Boltzmanngasse 9, 1090 Wien, Austria}

\emailAdd{andrea.campoleoni@umons.ac.be}
\emailAdd{lea.mele@umons.ac.be}

\abstract{Electric and magnetic Carrollian actions for fermions of arbitrary spin in any spacetime dimension are defined as inequivalent $c \to 0$ limits of the relativistic Fang–Fronsdal actions.}

\begin{document} 
\maketitle
\flushbottom


\section{Introduction}

Carrollian physics provides a natural framework to describe systems living on null hypersurfaces. Its name originates from the Carroll group \cite{Leblond, Gupta}, a contraction of the Poincaré group that plays, in this context, a role analogous to that of the latter in relativistic physics. The contraction is obtained by sending the speed of light $c$ to zero. Although such a limit might seem to be of doubtful physical relevance, it has found applications in rather diverse areas, including gravity \cite{Henneaux:1979vn, Duval:2014uva, Duval:2017els, Donnay:2019jiz, Oling:2024vmq, Fiorucci:2025twa, Fontanella:2025tbs} and condensed matter \cite{Casalbuoni:2021fel, Pena-Benitez:2021ipo, Bidussi:2021nmp, Figueroa-OFarrill:2023vbj, Ara:2024fbr}; see \cite{Bergshoeff:2022eog, Bagchi:2025vri, Ciambelli:2025unn} for reviews. In particular, the connection between Carrollian physics and null hypersurfaces has led to the conjecture that the gravitational dynamics of asymptotically flat spacetimes might admit a holographic description in terms of Carrollian field theories living on null infinity \cite{Bagchi:2016bcd, Ciambelli:2018wre, Donnay:2022aba, Bagchi:2022emh}; see \cite{Nguyen:2025zhg, Ruzziconi:2026bix} for reviews. 

These developments have motivated renewed interest in the study of Carroll-invariant field theories, whether formulated intrinsically \cite{Bagchi:2019clu, Gupta:2020dtl, Chen:2021xkw, Rivera-Betancour:2022lkc, Baiguera:2022lsw, deBoer:2023fnj, Bekaert:2024itn, Chen:2024voz, Cotler:2024xhb, Aggarwal:2025hji, Fredenhagen:2026pia, DArcy:2026hgl} or obtained from the \mbox{$c\to0$} limit of Lorentz-invariant ones \cite{Duval:2014uoa, Bagchi:2019xfx, Henneaux:2021yzg, deBoer:2021jej}. For our purposes, the latter viewpoint will be especially useful, and we shall focus on it.
As was first noticed in the case of electromagnetism \cite{Duval:2014uoa}, bosonic Lorentz-invariant actions typically admit two inequivalent limits \cite{Henneaux:2021yzg} (see also \cite{Barnich:2012rz} for an earlier example in two spacetime dimensions).\footnote{The two limits are generically inequivalent, but in some cases they are related by a remnant of the electric-magnetic duality of their relativistic parent \cite{Bekaert:2024itn, OConnor:2024rku}.} These limits are usually called electric and magnetic, following the terminology introduced in \cite{Duval:2014uoa, Henneaux:2021yzg} and also used for the Galilean \mbox{($c\to\infty$)} limit, where an analogous phenomenon occurs \cite{LeBellac:1973unm}. Electric and magnetic limits have in turn been defined for fields of spin $1/2$ using different approaches \cite{Koutrolikos:2023evq, Bergshoeff:2023vfd, Bagchi:2026lgk, Majumdar:2026iol}.\footnote{See also \cite{Bagchi:2022owq, Banerjee:2022ocj, Bagchi:2022eui, Zorba:2024jxb, Ekiz:2025hdn, Zheng:2025cuw, Bruce:2026yvw, Bulunur:2026yav, Bagchi:2026emg, Henneaux:2026dfc} for other works on fermionic Carrollian theories and \cite{Bergshoeff:2024ytq, Grumiller:2025rtm, Bagchi:2025vri} for reviews.} 
In this paper, we extend their characterisation to free massless fermions of arbitrary spin following \cite{Koutrolikos:2023evq, Bergshoeff:2023vfd}. To illustrate the key features of these two limits, we begin by recalling how they are defined for free scalars and spin-$1/2$ fermions.

We first consider the simpler electric limit. For a massive scalar field in a Minkowski spacetime of dimension $D = d+1$, the starting point is the Lorentz-invariant Klein--Gordon action
\begin{equation} \label{scalar}
S = \int dt \, d^{d}x \left( \frac{1}{2c^2}\, \pr_t \Phi \, \pr_t \Phi - \frac{1}{2}\, \pr_i \Phi \, \pr_i \Phi - \frac{(Mc)^2}{2}\, \Phi^2 \right) ,
\end{equation}
where, for later convenience, we follow the conventions of \cite{Henneaux:2021yzg} and assign the field a dimension such that the spatial-gradient contribution $(\partial_i \Phi)^2$ has the dimensions of an energy density (without any accompanying power of $c$).\footnote{Starting from the standard form of the Klein--Gordon action, eq.~\eqref{scalar} follows by setting $x^0 = c\, t$ and rescaling the field as $\Phi \to c^{-1/2} \Phi$. We focus here for simplicity on the limit of a free scalar field on a Minkowski background. On the other hand, the coupling of Carroll scalars to curved Carrollian backgrounds has also been studied in \cite{Rivera-Betancour:2022lkc, Baiguera:2022lsw}.}
The electric limit is then defined by sending $c \to 0$ after having rescaled the field as $\Phi = c\, \phi$ and the mass parameter as $M = c^{-2} m$. This leads to the electric Carroll-invariant action
\begin{equation} \label{electric_scalar}
S_{E} = \int dt \, d^{d}x \left(\frac{1}{2}\, \pr_t \phi \, \pr_t \phi - \frac{m^2}{2}\, \phi^2 \right) . 
\end{equation}

For a spin-$1/2$ field, one can follow the same strategy: assigning again a dimension to the field such that spatial gradients are not associated with any power of $c$ in the action, the Dirac action reads
\begin{equation} \label{dirac}
S = \int dt \, d^{d}x \left( \frac{1}{c}\, \bPsi \gamma^0 \pr_t \Psi + \bPsi \gamma^i \pr_i \Psi - Mc\, \bPsi \Psi \right) ,
\end{equation}
and admits the following $c \to 0$ limit with the redefinitions $\Psi = \sqrt{c}\, \psi$ and $M = c^{-2} m$ \cite{Bergshoeff:2017btm}:
\begin{equation} \label{electric_dirac}
S_{E} = \int dt \, d^{d}x \left( \bar{\psi} \gamma^0 \pr_t \psi - m\, \bar{\psi} \psi \right) .
\end{equation}
Let us stress that $\gamma^0$ denotes the usual gamma matrix, with the conventions of appendix~\ref{App:Conventions}. Indeed, in the previous manipulations we used the redefinition $x^0 = c\, t$, and we considered the Minkowski metric and the gamma matrices to be independent of $c$, as in \cite{Bergshoeff:2023vfd}.

We now move to the magnetic limit. To define it for the Klein-Gordon action \eqref{scalar}, it is convenient to cast the latter in Hamiltonian form:
\begin{equation} \label{H_scalar}
S = \int dt \, d^{d}x \left( \Pi \pr_t \Phi - \frac{c^2}{2}\, \Pi^2 - \frac{1}{2}\,  \pr_i \Phi \, \pr^i \Phi - \frac{(M c)^2}{2}\, \Phi^2 \right) .
\end{equation}
After this rewriting, one can directly take the limit $c \to 0$, provided one rescales the mass as $M=c^{-1}m$.
Identifying $\Pi = \pi$ and $\Phi = \phi$, this leads to the magnetic Carroll-invariant action \cite{Henneaux:2021yzg, deBoer:2021jej}
\be \label{magnetic_scalar}
S_{M} = \int dt\, d^dx \left( \pi \pr_t \phi - \frac{1}{2}\, \pr_i \phi \, \pr^i \phi - \frac{m^2}{2}\, \phi^2 \right) ,
\ee
whose equations of motion exhibit the triangular form
\be \label{eom_magnetic_scalar}
\partial_t \phi = 0 \, , \qquad \partial_t \pi = \left( \Delta - m^2 \right) \phi \, . 
\ee
Note that, in this limit, the conjugate momentum can no longer be eliminated algebraically through its equation of motion. If one rescales instead the fields in \eqref{H_scalar} as $\Pi = c^{-1} \pi$ and $\Phi = c\, \phi$ (so as to keep the kinetic term invariant), as well as the mass as $M = c^{-2} m$, the electric action \eqref{electric_scalar} is recovered by eliminating $\pi$ via its algebraic equation of motion. This scheme extends to the action of any bosonic massless field, since it essentially relies on the structure of the Hamiltonian form of a two-derivative Lorentz-invariant action. The Hamiltonian formulation therefore allows one to obtain both electric and magnetic limits, at least for all bosonic massless fields \cite{Henneaux:2021yzg}.

The Dirac action \eqref{dirac} is already of first order in derivatives, and the previous strategy cannot be applied verbatim. On the other hand, introducing the projectors 
\begin{equation} \label{projectors}
    P_\pm = \frac{1}{2} \left( \mathds{1} \pm i \, \gamma^0\right), 
\end{equation}
the Dirac equation $\left( \gamma^\mu \partial_\mu - M c \right) \Psi = 0$ splits into 
\begin{subequations} \label{magnetic_eom_dirac}
\begin{align}
i\, \partial_t \Psi_+ - c\, \gamma^k \partial_k \Psi_- + M c^2\, \Psi_+ & = 0 \, , \\
i\, \partial_t \Psi_- + c\, \gamma^k \partial_k \Psi_+ - M c^2\, \Psi_- & = 0 \, ,
\end{align}
\end{subequations}
where $\Psi_\pm := P_\pm \Psi$ and we took into account that $\gamma^0 \Psi_{\pm} = \mp\, i\, \Psi_{\pm}$. Introducing the rescalings $\Psi_+ = c^{-1/2}\, \psi_+$, $\Psi_- = c^{1/2} \,\psi_-$ and \mbox{$M = c^{-2} m$}, we obtain in the $c \rightarrow 0$ limit \cite{Bagchi:2019xfx, Bergshoeff:2023vfd}
\begin{equation} \label{eom_magnetic_dirac}
     \left( i\, \partial_t + m \right) \psi_+ = 0 \, , \qquad \left( i\, \partial_t - m \right) \psi_- + \gamma^k \partial_k \psi_+ = 0 \, .
\end{equation}
These equations of motion have a triangular structure similar to that of \eqref{eom_magnetic_scalar}, and one can use them to define the magnetic theory.\footnote{The fields $\psi_{\pm}$ form a reducible but indecomposable representation of the Carroll group \cite{Bergshoeff:2023vfd}, again in analogy with the bosonic fields entering \eqref{eom_magnetic_scalar} (see, e.g., \cite{deBoer:2021jej, Bekaert:2024itn, Chen:2024voz}).} Obtaining them from an action principle is however subtle. Various approaches have been developed \cite{Koutrolikos:2023evq, Bergshoeff:2023vfd, Bagchi:2026lgk, Majumdar:2026iol}, and we now review the one proposed in \cite{Koutrolikos:2023evq, Bergshoeff:2023vfd} and that we shall extend in the following to massless fermions of arbitrary spin.

Even if the Dirac action \eqref{dirac}, being of first order, is already in Hamiltonian form, one can introduce a conjugate momentum for each field variable as
\be
\Pi := \frac{\partial \cL}{\partial \dot{\Psi}} = \frac{1}{2c}\, \bar{\Psi} \gamma^0 \, . 
\ee
The action \eqref{dirac} can therefore be rewritten as 
\begin{equation} \label{H_dirac}
S =  \int dt \, d^{d}x \left\{ \Pi \dot{\Psi} + \frac{1}{2}\, \bar{\Psi} \gamma^i \pr_i \Psi + Mc^2\, \Pi \, \gamma^0 \Psi + \left( \Pi - \frac{1}{2c}\, \bar{\Psi} \gamma^0 \right) \lambda \right\} + \textrm{h.c.} \, ,
\end{equation}
where a dot denotes a time derivative and $\lambda$ is a Lagrange multiplier enforcing the second-class constraint
\be \label{second-class_dirac}
\Pi \approx \frac{1}{2c}\, \bar{\Psi} \gamma^0 \, .
\ee
The equations of motion then show that on shell $\lambda$ is determined by the spatial derivatives of $\Psi$ and therefore does not propagate.
Rescaling the mass parameter as $M = c^{-2} m$ and the Lagrange multiplier as $\lambda = c^{\,\varepsilon} \tilde{\lambda}$ with $\varepsilon > 1$, in the limit $c \to 0$ the terms enforcing the second-class constraint disappear, and one obtains the action
\begin{equation} \label{magnetic_dirac}
S_{M} =  \int dt \, d^{d}x \left\{ \pi \dot{\psi} +  \dot{\bar{\psi}} \bar{\pi}+ \bar{\psi} \gamma^i \pr_i \psi + m \left( \pi \gamma^0\psi - \bar{\psi} \gamma^0 \bar{\pi} \right) \right\} ,
\end{equation}
where we also introduced $\Pi = \pi$ and $\Psi = \psi$ to keep the notation uniform with that adopted for the electric limit. Moreover, $\bar{\pi} = i \gamma^0 \pi^\dagger$.
Its equations of motion have a triangular form similar to that of \eqref{eom_magnetic_dirac},
\be \label{magnetic_1}
\left( \partial_t  + m \,  \gamma^0 \right) \psi = 0 \, , \qquad
\left(\partial_t  + m \, \gamma^0 \right) \bar{\pi}- \gamma^k \partial_k \psi = 0 \, ,
\ee
but they contain more spinor fields with respect to the latter. However, these can be recovered by projecting eqs.~\eqref{magnetic_1} using $P_\pm$ and then performing the truncation
\be
P_- \bar{\pi} := - \,\gamma^0 \psi_- \,, \qquad 
P_+ \psi := \psi_+ \,, \qquad  
P_+ \bar{\pi} = 0 \,, \qquad
P_- \psi = 0 \,.
\ee
Note that this truncation cannot be implemented directly into the action \eqref{magnetic_dirac} because it would lead to its complete cancellation. In analogy with the bosonic case, in this setup the electric theory can be recovered by rescaling the fields as $\Pi = c^{-1/2} \pi$, $\Psi = c^{1/2} \psi$ (so as to keep the kinetic term invariant), and the Lagrange multiplier as $\lambda = c^{1/2} \tilde{\lambda}$ while keeping the same rescaling for the mass. In this way, the second-class constraints are preserved and one can use them to eliminate algebraically the conjugate momentum to recover the action \eqref{electric_dirac}.

At the level of the action, it is therefore natural to define the magnetic Carrollian limit by removing the second-class constraint relating $\Pi$ and $\bar{\Psi}$, thereby doubling the number of independent spinor fields.\footnote{An additional spinor also appears in other approaches, see, e.g., the other action discussed in \cite{Bergshoeff:2023vfd}.} The rest of the action remains unchanged in the limit, up to possible substitutions of $\bar{\Psi}$ with $\Pi$, that before the limit can always be performed either using the second-class constraint \eqref{second-class_dirac} or, equivalently, by redefining the Lagrange multiplier $\lambda$ \cite{Bergshoeff:2023vfd}. In particular, one can also consider the $c \to 0$ limit of the following rewriting of eq.~\eqref{H_dirac}, exhibiting a different form of the mass term:
\begin{equation} \label{H_dirac_2}
S =  \int dt \, d^{d}x \left\{ \Pi \dot{\Psi} + \frac{1}{2}\, \bar{\Psi} \gamma^i \pr_i \Psi - \frac{Mc}{2}\, \bar{\Psi} \Psi + \left( \Pi - \frac{1}{2c}\, \bar{\Psi} \gamma^0 \right) \lambda \right\} + \textrm{h.c.}
\end{equation}
Before the limit, the two actions are equivalent. After taking the $c \to 0$ limit with the same rescalings for all fields as before but imposing instead $M = c^{-1} m$, this action gives the equations of motion
\be \label{magnetic_2}
\partial_t \psi = 0 \, , \qquad
\partial_t \bar{\pi} = \left( \gamma^k \partial_k - m \right) \psi \, .
\ee
Introducing
\be \label{redef}
\chi = \left( \gamma^k \partial_k + m \right) \bar{\pi} \,,
\ee
these equations imply 
\be
\partial_t \psi = 0 \, , \qquad
\partial_t \chi = \left( \Delta - m^2 \right) \psi \, .
\ee
The components of the two spinors $\chi$ and $\psi$ therefore satisfy the bosonic magnetic equations of motion \eqref{eom_magnetic_scalar}, just as the Dirac equation implies the Klein--Gordon equation for each spinor component. In this sense, the doubling of independent spinor fields in the Carroll-invariant action \eqref{magnetic_dirac} naturally mirrors the fact that the conjugate momentum of the scalar field can no longer be eliminated algebraically through its equation of motion. This analogy suggests that the two-spinor unprojected equations of motion \eqref{magnetic_1} and \eqref{magnetic_2} should also be regarded as natural candidates for the magnetic theory. For instance, in \cite{Koutrolikos:2023evq} they have been considered as the starting point to build a Carrollian analogue of the Wess-Zumino model involving magnetic scalars.

Let us stress that in the massless case, which will be our main focus in the following, the actions \eqref{H_dirac} and \eqref{H_dirac_2} coincide. We have discussed the $c \to 0$ limits of a massive Dirac field in order to illustrate the subtleties introduced by the mass term, which appears to allow for inequivalent magnetic limits. In the massless limit, by contrast, this ambiguity disappears: one can consider either the non-Lagrangian projected equations \eqref{eom_magnetic_dirac} or their two-spinor triangular Lagrangian extension \eqref{magnetic_2}. Moreover, in the massless case different rewritings of the spatial-derivative term in the action simply lead to equations related by field redefinitions \cite{Bergshoeff:2023vfd}.

Motivated by these considerations, in the following we introduce conjugate momenta for free massless fermionic fields of arbitrary spin and define electric and magnetic limits for the corresponding actions. In the electric case, we keep the second-class constraints, while spatial gradients are lost in the Carrollian limit. In the magnetic case, instead, we remove the second-class constraints in the $c \to 0$ limit, while retaining all other terms in the action. As we shall see, particular care is needed to preserve the first-class constraints that appear starting from spin-$3/2$ fermions. In the magnetic case, we also show that the resulting equations of motion can be truncated to those obtained by taking a limit similar to \eqref{magnetic_eom_dirac} directly at the level of the Fang--Fronsdal equations of motion. 
We first carry out this analysis for spin-$3/2$ fields, which are relevant for supergravity theories, and then discuss explicitly the spin-$5/2$ example before moving to the arbitrary-spin case. Our conventions are detailed in appendix~\ref{App:Conventions}, while appendix~\ref{app:H} presents the Fang--Fronsdal action \cite{Fang:1978wz} for fermions of arbitrary spin in Hamiltonian form, following \cite{Campoleoni:2017vds} but including additional details not given there, such as the explicit expressions for the symplectic form and the Hamiltonian.

\section{Spin 3/2} \label{sec:3/2}

We describe a spin-$3/2$ field using the vector-spinor $\Psi_\mu^\alpha$ and the Lorentz-invariant Rarita--Schwinger action
\begin{equation} \label{rarita-schwinger}
    S =  c^{-1} \int d^Dx\, \bar{\Psi}_\mu \gamma^{\mu\nu\rho}\partial_\nu \Psi_\rho \,,
\end{equation}
where here and in the following spinorial indices are omitted and $\gamma^{\mu\nu\rho}=\gamma^{\left[\mu\right.} \gamma^\nu \gamma^{\left.\rho\right]}$ (our conventions are spelt out in appendix~\ref{App:Conventions}). As in the Introduction, we assign a dimension to the fields such that spatial gradients are not associated with any power of $c$ in the action. The Rarita-Schwinger action is invariant under the gauge transformation
\begin{equation}
    \Psi_\mu(t,\mathbf{x}) \rightarrow \Psi_\mu(t,\mathbf{x}) + \partial_\mu \epsilon (t,\mathbf{x}) \,, \label{gaugetransfospin3/2}
\end{equation}
with $\epsilon(t,\mathbf{x})$ being a Dirac spinor. 

\paragraph{Action in Hamiltonian form.}

Separating temporal and spatial components of the field, eq.~\eqref{rarita-schwinger} reads
\be \label{rarita-schwinger_2}
S = \int dt\, d^d x \left\{ \frac{1}{2c}\left( \Psi^\dag_k \omega^{kl} \dot{\Psi}_l - \dot{\Psi}^\dag_k \omega^{kl} \Psi_l  \right) - \Psi^\dag_0 \mathcal{F}- \mathcal{F}^\dag \Psi_0 - \mathcal{H}\right\} .
\ee
This action is already in Hamiltonian form, with $\Psi_k$ and $\Psi^\dagger_k$ playing the role of canonical variables.\footnote{For simplicity, in the Introduction we wrote the kinetic terms using the Dirac conjugate as in the usual covariant rewriting of the action, while from now on we shall follow the conventions of, e.g., \cite{Campoleoni:2017vds} for the kinetic terms.} The symplectic two-form is
\be
\omega^{kl} = i\, \gamma^{kl} \, ,
\ee
while the Hamiltonian reads
\begin{equation}
    \mathcal{H} = -\bar{\Psi}_k\gamma^{klm}\partial_l \Psi_m \, .
\end{equation}
The temporal component $\Psi_0$ of the vector-spinor is a Lagrange multiplier enforcing the first-class constraint 
\begin{equation} \label{1st-class_3/2}
    \mathcal{F} := i\, \gamma^{kl}\partial_k \Psi_l \approx 0 \, ,
\end{equation}
which generates the gauge transformations \eqref{gaugetransfospin3/2}.

As in the spin-$1/2$ case, one can nevertheless introduce a conjugate momentum for each field variable to eliminate the factor $c^{-1}$ in front of the kinetic term. The action can then be cast in the form
\begin{equation}
\begin{aligned}
      S = \int dt\, d^dx & \bigg\{\, \Pi^k \dot{\Psi}_k  + \dot{\Psi}_k^\dag\Pi^{\dag k} - \Psi_0^\dag \mathcal{F} - \mathcal{F}^\dag \Psi_0 - \mathcal{H} \\ 
    & + \left( \Pi^k - \frac{1}{2c}\, \Psi^\dag_l \omega^{lk} \right) \lambda_k + \lambda^\dag_k \left( \Pi^{\dag k} + \frac{1}{2c}\, \omega^{kl}\Psi_l \right) \bigg\} \,.  \label{Hamiltonian form action 3/2}
\end{aligned}
\end{equation}
The terms in the first line of eq.~\eqref{Hamiltonian form action 3/2} directly come from the rewriting \eqref{rarita-schwinger_2} of the Rarita--Schwinger action. In the second line, we added the Lagrange multiplier $\lambda_l$ enforcing the second-class constraint
\begin{equation} \label{2nd-class_3/2}
    \mathcal{C}^k := 
    \Pi^k - \frac{1}{2c}\, \Psi^\dag_l \omega^{lk}  \approx 0 \,,
\end{equation}
and its Hermitian conjugate. 
Note that, after introducing the conjugate momenta $\Pi^k$ and $\Pi^{\dag k}$, the action depends on the speed of light $c$ only via the second-class constraints. 
We now discuss electric and magnetic limits along the lines presented in the Introduction.

\paragraph{Electric limit.}

In analogy with the spin-$1/2$ case, the electric limit is obtained by rescaling the canonical variables as 
\begin{equation} \label{electric_rescale_3/2_1}
    \Psi_k = \sqrt{c} \, \psi_k \,, \qquad \Pi^k = \frac{1}{\sqrt{c}}\, \pi^k \,,
\end{equation}
as well as the Lagrange multipliers as
 \begin{equation} \label{electric_rescale_3/2_2}
     \Psi_0 =  \frac{1}{\sqrt{c}}\, \psi_0 \,, \qquad \lambda_k = \sqrt{c}\, \tilde{\lambda}_k \,.
 \end{equation}
The rescalings of the canonical variables leave the kinetic term invariant while removing the relative factor of $c$ in the second-class constraint, thus allowing one to preserve it in the $c\to0$ limit. The Lagrange multipliers are then rescaled so as to retain all constraints in the limit.
Sending $c \rightarrow 0$, one obtains the action
\begin{equation}
\begin{aligned}
      S_{E} = \int dt\, d^dx &\bigg\{\, \pi^k \dot{\psi}_k  + \dot{\psi}_k^\dag\pi^{\dag k} - \psi_0^\dag \mathcal{F} - \mathcal{F}^\dag \psi_0  \\ 
    & + \left( \pi^k - \frac{1}{2}\, \psi^\dag_l \omega^{lk} \right) \tilde{\lambda}
    _k + \tilde{\lambda}^\dag_k \left( \pi^{\dag k} + \frac{1}{2}\, \omega^{kl}\psi_l \right)   \bigg\} \,, \label{elecaction3/2}
\end{aligned}
\end{equation}
where we stress that $\cH = 0$, while 
\be \label{Carroll_first-class_3/2}
    \mathcal{F}=  i\, \gamma^{kl}\partial_k \psi_l \, ,
\ee
and $\omega^{kl}= i\, \gamma^{kl}$ as before the limit.

The conjugate momenta can be eliminated enforcing the second-class constraints and one obtains the action
\be
S_E = i \int dt\, d^dx \left\{ \frac{1}{2}\, \psi^\dag_k \gamma^{kl} \dot{\psi}_l -  \psi^\dag_0 \gamma^{kl} \partial_k \psi_l \right\} + \textrm{h.c.}
\ee
that can also be obtained directly from the action \eqref{rarita-schwinger} by taking the $c \to 0$ limit with the rescalings $\Psi_k = c^{1/2}\, \psi_k$ and $ \Psi_0 = c^{-1/2}\, \psi_0$. This action is invariant under the gauge transformations
\be
\delta \psi_k = \partial_k \epsilon \, , \qquad
\delta \psi_0 = \partial_t \epsilon \, ,
\ee
that can also be obtained from \eqref{gaugetransfospin3/2} by rescaling the gauge parameter as $\epsilon \to c^{1/2} \epsilon$. The gauge transformations of the canonical variables are generated by the first-class constraint \eqref{Carroll_first-class_3/2}. The transformation of the Lagrange multiplier is instead fixed by requiring that the Hamiltonian equations of motion retain the same form under time evolution. Indeed, since $\cH=0$, the time evolution of the canonical variables takes the form of a gauge transformation.

\paragraph{Magnetic limit.}

Again in analogy with the spin-$1/2$ case, the magnetic limit can be defined by sending $c\to0$ without any rescaling of the canonical variables. The divergence in the second-class constraint (and in its hermitian conjugate) can then be removed by rescaling the Lagrange multiplier that enforces it so that the second line of eq.~\eqref{Hamiltonian form action 3/2} drops out in the limit. In this way, one can retain a non-vanishing Hamiltonian in the Carrollian limit at the price of losing the second-class constraints \eqref{2nd-class_3/2}. Compared with the spin-$1/2$ example, however, one also has to discuss how to handle the first-class constraint \eqref{1st-class_3/2} and the Lagrange multiplier $\Psi_0$. 
Since we wish the second line of eq.~\eqref{Hamiltonian form action 3/2} to vanish in the limit, it is convenient to rewrite the action so that this term is separately gauge invariant. This can be achieved through the field redefinition $\lambda_k \to \lambda_k - c\, \partial_k \Psi_0$, which introduces a term proportional to the rewriting of the first-class constraint in terms of the conjugate momenta. However, this term is proportional to $c$ and therefore disappears in the $c\to0$ limit. To obtain a Carrollian limit in which both $\psi_k$ and $\pi^k$ retain non-trivial gauge transformations, we introduce instead the field redefinitions 
\begin{equation} \label{redef_multipliers_3/2}
    \Psi_0 = \chi + c^{-1} \zeta \,, \qquad
    \lambda_k = \tilde{\lambda}_k - 2\,\partial_k \zeta \,.
\end{equation}
This leads to an equivalent rewriting of the action \eqref{Hamiltonian form action 3/2},
\begin{equation} \label{S_H_new_3/2}
\begin{split}
    S = \int dt\, d^d x & \bigg\{\, \Pi^k \dot{\Psi}_k + \dot{\Psi}^{\dag k}\Pi^\dag_k - \chi^\dag \mathcal{F} - \zeta^\dag \mathcal{G} - \mathcal{F}^\dag \chi - \mathcal{G}^\dag \zeta - \mathcal{H}  \\
    &  + \left( \Pi^k - \frac{1}{2c}\, \Psi^\dag_l \omega^{lk} \right) \tilde{\lambda}_k + \tilde{\lambda}^\dag_k \left( \Pi^{\dag k} + \frac{1}{2c}\, \omega^{kl}\Psi_l \right) \bigg\} \,,
\end{split}
\end{equation}
with
\be \label{Carroll_1st-class_3/2}
    \mathcal{F}=  i\, \gamma^{kl}\partial_k \Psi_l \, , \qquad
    \mathcal{G} = - 2\, \partial_k \Pi^{\dag k} \,.
\ee
The constraint $\mathcal{G}$ can also be obtained by rewriting $c^{-1}\mathcal{F}$ in terms of $\Pi_k^\dagger$, using the expression for $\Psi_k$ that follows from solving the second-class constraint \eqref{2nd-class_3/2}. The form of the field redefinition \eqref{redef_multipliers_3/2} can then be deduced by noticing that
\be \label{relation_constraints_3/2}
2\,\partial_k \mathcal{C}^k = c^{-1} \mathcal{F}^\dag - \mathcal{G}^\dag\, ,
\ee
where $\mathcal{C}^k$ is the second-class constraint \eqref{2nd-class_3/2}. At finite $c$, this rewriting of the action is fully equivalent to \eqref{Hamiltonian form action 3/2}. Indeed, the new parametrisation introduces the redundancy
\be
\delta \chi = \alpha \, , \qquad
\delta \zeta = - c\, \alpha \, , \qquad
\delta \tilde{\lambda}_k = -2c\,\partial_k \alpha \, , 
\ee
so that the enlarged set of constraints is reducible and imposes no additional independent conditions, as is also reflected in the relation \eqref{relation_constraints_3/2}.

The magnetic limit is then obtained by sending $c \to 0$ after rescaling $\tilde{\lambda}_k = c^{\,\varepsilon} \hat{\lambda}_k$ with $\varepsilon > 1$. This gives the action
\begin{equation}
    S_M = \int dt\, d^d x \left( \pi^k \dot{\psi}_k + \dot{\psi}^{\dag k}\pi^\dag_k -  \chi^\dag \mathcal{F} - \zeta^\dag \mathcal{G} - \mathcal{F}^\dag \chi - \mathcal{G}^\dag \zeta - \mathcal{H}  \right) , \label{magnaction_spin3/2}
\end{equation}
where we stress that we kept all terms with spatial derivatives, i.e., the first-class constraints and the Hamiltonian $\cH$.
The constraints $\mathcal{F}$ and $\mathcal{G}$ are of first class, as they satisfy the Poisson algebra
\begin{equation}
\big\{ \mathcal{F}, \mathcal{F}^\dag \big\} = 0\, , \qquad
\big\{ \mathcal{F}, \mathcal{G}^\dag \big\} = 0 \, , \qquad
\big\{ \mathcal{G}, \mathcal{F}^\dag \big\} = 0 \, , \qquad 
\big\{ \mathcal{G}, \mathcal{G}^\dag \big\} = 0 \, .
\end{equation}
Moreover, there are no secondary constraints even after the Carrollian limit, since
\begin{equation} \label{secondary_3/2}
\big\{ \mathcal{F}, \mathcal{H} \big\} = 0\, , \qquad
\big\{ \mathcal{G}, \mathcal{H} \big\} = 0 \, .
\end{equation}
The Hamiltonian of the magnetic action only depends on half of the new canonical variables, so that one directly obtains
\be
\big\{ \cH(x) , \cH(x') \big\} = 0 \, .
\ee
As discussed in \cite{Henneaux:2021yzg}, this condition, together with the scalar behaviour under spatial rotations and translations, suffices to guarantee the invariance of the magnetic action \eqref{magnaction_spin3/2} under the Carroll group. Moreover, in this case the Hamiltonian is gauge invariant up to a boundary term and this guarantees that its Poisson bracket with the first-class constraints weakly vanishes (as explicitly shown in \eqref{secondary_3/2}), which is another necessary condition for the Carroll-invariance of the action for gauge theories.

With this procedure, we doubled the number of independent spinor fields by removing the second-class constraints as in the spin-$1/2$ case. On the other hand, each spinor kept the same amount of gauge symmetry as before the limit, thus effectively doubling also the number of independent gauge parameters. Indeed, the first-class constraints \eqref{Carroll_1st-class_3/2} generate the gauge transformations
\be
\delta \psi_k = \partial_k \epsilon \, , \qquad
\delta \pi^k = \frac{1}{2}\,\partial_l \rho^\dag \omega^{lk} \, ,
\ee
which leave the action \eqref{magnaction_spin3/2} invariant when combined with the following transformations for the Lagrange multipliers
\be
\delta \chi = \frac{1}{2}\,\partial_t \rho \, , \qquad
\delta \zeta = \frac{1}{2}\,\partial_t \epsilon \, .
\ee

\paragraph{Minimal magnetic equations of motion.}

The equations of motion of the Rarita--Schwinger action \eqref{rarita-schwinger}, 
\begin{equation}
    \gamma^{\mu \nu \rho} \partial_\nu \Psi_{\rho}=0 \, ,
\end{equation}
can be used to define a magnetic limit similar to that we defined in eq.~\eqref{eom_magnetic_dirac} for a spin-$1/2$ field. Acting on them with the projection operators $P_\pm$ defined in \eqref{projectors} and implementing the rescalings
\begin{equation}
\Psi_{j\pm} = c^{\mp\frac{1}{2}} \, \psi_{j\pm} \, , \qquad
\Psi_{0\pm} = c^{-1\mp\frac{1}{2}} \, \psi_{0\pm} \, ,
\end{equation}
in the $c \rightarrow 0$ limit we get 
\begin{subequations} \label{eom_magnetic_3/2}
\begin{align}
    & \gamma^{ij0}\partial_t\psi_{j+}-\gamma^{ij0}\partial_j \psi_{0+}=0 \, , \\[5pt]
    & \gamma^{ij0}\partial_t \psi_{j-}-\gamma^{ij0}\partial_j \psi_{0-} - \gamma^{ijk}\partial_j \psi_{k+}=0 \, , \\[5pt]
    & \gamma^{ij0}\partial_i \psi_{j \pm} = 0 \, .
\end{align}
\end{subequations}

The equations of motion of the action \eqref{magnaction_spin3/2} are instead 
\begin{subequations}
\begin{align}
    &\partial_t \psi_k - 2\,\partial_k \zeta = 0 \, , \\[5pt]
    & \partial_t \pi^{\dag k} + i\, \gamma^{kl}\partial_l \chi - i\,  \gamma^0 \gamma^{klm} \partial_l \psi_m = 0 \, ,
\end{align}
\end{subequations}
together with the constraints
\be
    \partial_l \pi^{\dag l}=0 \, , \qquad i \, \gamma^{kl} \partial_k \psi_l = 0 \, .
\ee
These equations can be equivalently rewritten as 
\begin{subequations}
\begin{align}
     &\gamma^{ij0}\partial_t \psi_{j\pm} - 2\, \gamma^{ij0} \partial_j \zeta_\pm = 0 \, , \\[5pt]
    & i\, \gamma^0 \partial_t \pi^{\dag i}_\pm + \gamma^{0ki}\partial_k \chi_\pm - \gamma^{ilm} \partial_l \psi_{m\mp} = 0 \, , \\[5pt]
    & \gamma^0 \partial_l \pi^{\dag l}_\pm=0, \quad i\, \gamma^{0kl} \partial_k \psi_{l \pm} = 0 \, ,
\end{align}
\end{subequations}
where we already acted on them with the projectors $P_\pm$.
Imposing the truncation
\begin{subequations}
\begin{alignat}{5}
P_+ \psi_{j} & = \psi_{j+} \, , & \qquad P_- \psi_j & = 0 \, , & \qquad 
P_+ \pi^{\dag j} & = 0 \, , & \qquad P_- \pi^{\dag j} & = - i\, \gamma^{jk}\psi_{k-}  \, , \\
P_+ \chi & = 0 \, , & \qquad P_- \chi & = \psi_{0-} \, , & \qquad 
P_+ \zeta & = \frac{1}{2}\,\psi_{0+} \, , & \qquad P_- \zeta & = 0 \, ,
\end{alignat}
\end{subequations}
we recover the projected magnetic equations \eqref{eom_magnetic_3/2}.

\section{Spin 5/2} \label{sec:5/2}

We follow the approach of Fang and Fronsdal to describe fermions of arbitrary spin \cite{Fang:1978wz}. Before moving to the generic case, we first discuss explicitly the example of a spin-$5/2$ field, which already exhibits most of the key features of higher-spin fermions. We describe such a field using a spinor-tensor $\Psi_{\mu\nu}$, which is symmetric in its spacetime indices, $\Psi_{\mu\nu}= \Psi_{\nu\mu}$. We also define $\Psi' := \eta^{\mu\nu} \Psi_{\mu\nu}$. The Fang-Fronsdal action reads
\begin{equation}
\begin{aligned}
    S = c^{-1} \int d^Dx & \bigg\{\, \bPsi^{\mu\nu} \gamma^\sigma \partial_\sigma \Psi_{\mu\nu} + 2\, \bPsi_{\mu\nu} \gamma^\nu \gamma^\lambda \gamma^\rho \partial_\lambda \Psi_\rho{}^\mu - \frac{1}{2}\, \bPsi' \gamma^\sigma \partial_\sigma \Psi' \\
    & + \bPsi_{\mu\nu} \gamma^\mu \partial^\nu \Psi' + \bPsi' \gamma^\mu \partial^\nu  \Psi_{\mu\nu} - 2 \left(\bPsi^{\mu\nu} \gamma^\rho \partial_\mu  \Psi_{\rho \nu} +\bPsi_{\mu\nu}\gamma^\mu \partial_\rho \Psi^{\rho \nu} \right) \bigg\} \,.
    \label{spin52action}
\end{aligned}
    \end{equation}
This action is invariant under the gauge transformations
\begin{equation} \label{gauge_5/2}
    \delta \Psi_{\mu\nu} = \partial_\mu \epsilon_\nu + \partial_\nu \epsilon_\mu \,, 
\end{equation}
with the spinor-vector $\epsilon_\mu$ obeying the algebraic constraint $\gamma^\mu \epsilon_\mu = 0$.

\paragraph{Action in Hamiltonian form.}

Since the action \eqref{spin52action} is of first order in derivatives, casting it in Hamiltonian form essentially amounts to distinguishing the canonical variables from the Lagrange multipliers that enforce the first-class constraints generating the gauge symmetry. This analysis has been performed in \cite{Aragone:1979hw} for fields of arbitrary half-integer spin and revisited in the spin-$5/2$ case in \cite{Borde:1981gh} and more recently in \cite{Bunster:2014fca, Campoleoni:2017vds}. Hamiltonian actions inspired by the frame formulation of general relativity have also been built in both Minkowski \cite{Aragone:1980rk} and (A)dS backgrounds \cite{Vasiliev:1987hv}. Here we follow \cite{Campoleoni:2017vds}, adapting the conventions to those of \cite{Bergshoeff:2023vfd} (see appendix~\ref{App:Conventions} for the details).

We begin by recalling that time derivatives of the gauge parameters can only appear in the gauge variations of Lagrange multipliers. Thanks to this criterion, we can identify the dynamical variables with the spatial components $\Psi_{ij}$ and with the combination
\begin{equation} \label{def_xi_5/2}
     \Xi = \Psi_{00} - 2\, \gamma^0 \gamma^k \Psi_{0k} \,.
\end{equation}
With this field redefinition, the gauge transformations \eqref{gauge_5/2} read
\be
    \delta \Psi_{kl} =\partial_k \epsilon_l + \partial_l \epsilon_k \, , \qquad 
    \delta \Xi = -2\, \gamma^k \gamma^l \partial_k \epsilon_l \, , \qquad
    \delta \Psi_{0k} = c^{-1}\,\partial_t \epsilon_k + \gamma^0 \gamma^l \partial_k \epsilon_l \, ,
    \label{gaugetransfospin5/2}
\ee
where we used the constraint $\gamma^\mu \epsilon_\mu = 0$ to replace $\epsilon_0 = \gamma^0 \gamma^k \epsilon_k$ in the variation of $\Xi$. This identifies $\Psi_{0k}$ as a Lagrange multiplier. Accordingly, the action \eqref{spin52action} can be cast in Hamiltonian form as
\begin{equation} \label{S_H_5/2}
\begin{split}
    S = \int dt\, d^d x & \bigg\{ \left( \Pi^A\dot{\Psi}_A + \dot{\Psi}^\dag_A \Pi^{\dag A}\right) - \Psi^\dag_{0k}\mathcal{F}^k-\mathcal{F}^{\dag k} \Psi_{0k}-\mathcal{H} \\
    & + \left( \Pi^A - \frac{1}{2c}\, \Psi^\dag_B \omega^{BA} \right)\lambda_A + \lambda_A^\dag\left( \Pi^{\dag A} + \frac{1}{2c}\, \omega^{AB} \Psi_B \right)  \bigg\} \,, 
    \end{split}
\end{equation}
where
\begin{equation}
\Psi_A=\left(\begin{array}{c}\Psi_{k l} \\ \Xi \end{array}\right) , \quad
\Pi^A=\left(\begin{array}{c}\Pi^{k l} \\ \Theta \end{array}\right) , \quad
\omega^{A B}=\left(\begin{array}{cc}
\omega^{k l \mid m n} & \omega^{k l \mid \bullet} \\
\omega^{\bullet \mid m n} & \omega^{\bullet \mid \bullet}
\end{array}\right) ,
\end{equation}
with
\be
\Pi^{kl} := \frac{\partial \cL}{\partial \dot{\Psi}_{kl}} = \frac{1}{2c} \left( \Psi_{mn}^\dag \omega^{mn \mid kl} + \Xi^\dag \omega^{\bullet \mid kl} \right) , \quad
\Theta := \frac{\partial \cL}{\partial \dot{\Xi}} = \frac{1}{2c} \left( \Psi_{kl}^\dag \omega^{kl \mid \bullet} + \Xi^\dag \omega^{\bullet \mid \bullet} \right) .
\ee
Compared to \cite{Campoleoni:2017vds}, we already introduced a conjugate momentum for each field variable as in section~\ref{sec:3/2}, as well as Lagrange multipliers enforcing the corresponding second-class constraints. As in all previous examples, this moves the factor $c^{-1}$ from the kinetic term to the second line of \eqref{S_H_5/2}.
The components of the symplectic two-form read
\begin{subequations} \label{omega_5/2}
\begin{align}
\omega^{k l \mid m n} & = i \left( -g^{k(m} g^{n) l} +2\, \gamma^{(k} g^{l)(m} \gamma^{n)} +\frac{1}{2}\, g^{k l} g^{m n}\right) , \\
\omega^{k l \mid \bullet} & =\omega^{\bullet \mid k l}=\frac{i}{2}\,  g^{k l} \,, \\ \omega^{\bullet \mid \bullet} & = -\frac{i}{2} \,,
\end{align}
\end{subequations}
while the Hamiltonian and the first-class constraints read, respectively,
\begin{equation} \label{H_5/2}
    \begin{aligned}
        \mathcal{H} = & -\bPsi^{k l} \gamma^m \partial_m \Psi_{k l} + \frac{3}{2}\, \bar{\Xi} \gamma^k \partial_k \Xi - 2\, \bPsi_{k l} \gamma^l \gamma^m \gamma^n \partial_{m} \Psi_n{}^k - \bar{\Psi}_{k l} \gamma^k \partial^l \Psi_m{}^m \\
        & -\bar{\Psi}_k{}^k \gamma^l \partial_m \Psi_{l}{}^{m} +2\, \bPsi^{k l} \gamma^m \partial_k \Psi_{m l}+2\, \bar{\Psi}_{k l} \gamma^k \partial_m \Psi^{m l}+\frac{1}{2}\, \bar{\Psi}_k{}^k \gamma^l \partial_l \Psi_m{}^m  \\
&  -\frac{1}{2}\, \bar{\Xi} \gamma^k \partial_k \Psi_l{}^l -\frac{1}{2}\, \bPsi_k{}^k \gamma^l\partial_l \Xi+\bPsi_k{}^l\gamma^k \partial_l \Xi +\bar{\Xi} \gamma^k \partial_l \Psi_k{}^l \,, 
\end{aligned}
\end{equation}
and
\begin{equation} \label{F_5/2}
    \begin{split}
\mathcal{F}_k := i & \Big[\,\partial_k \Xi - 2\, \partial_l \Psi_k{}^l+\partial_k \Psi_l{}^l + 2\, \gamma^l \gamma^m \partial_l \Psi_{km} \\
& + \gamma_k \left( \gamma^l \partial_l \Xi 
 +2\, \gamma^l \partial_m \Psi_l{}^m - \gamma^l \partial_l \Psi_m{}^m \right)\Big] \approx 0 \,.
 \end{split}
\end{equation}
Note that no factors of $c$ appear in eqs.~\eqref{omega_5/2}--\eqref{F_5/2}. We now use this rewriting of the action to define electric and magnetic limits along the lines already detailed for the spin-$3/2$ case.

\paragraph{Electric limit.} The strategy for defining the electric limit is the same as in the previous examples: we rescale the dynamical variables in a way that leaves the kinetic term invariant while removing the relative factor of $c$ in the second-class constraints. The presence of the field $\Xi$ in addition to $\Psi_{ij}$ does not cause any additional difficulty and eqs.~\eqref{electric_rescale_3/2_1} and \eqref{electric_rescale_3/2_2} are generalised by
\begin{equation}
    \Psi_A = \sqrt{c} \, \psi_A \,, \qquad 
    \Pi^A = \frac{1}{\sqrt{c}}\, \pi^A \,, \qquad
    \Psi_{0k} =  \frac{1}{\sqrt{c}}\, \psi_{0k} \,, \qquad 
    \lambda_A = \sqrt{c}\, \tilde{\lambda}_A \,.
 \end{equation}
Sending $c \rightarrow 0$, one obtains the action
\begin{equation} \label{electricaction52}
\begin{aligned}
    S_E = \int dt\, d^d x & \bigg\{ \left( \pi^A\dot{\psi}_A + \dot{\psi}^\dag_A \pi^{\dag A}\right) - \psi^\dag_{0k}\mathcal{F}^k-\mathcal{F}^{\dag k} \psi_{0k} \\
    & + \left( \pi^A - \frac{1}{2}\, \psi^\dag_B \omega^{BA} \right) \tilde{\lambda}_A + \tilde{\lambda}_A^\dag\left( \pi^{\dag A} + \frac{1}{2}\, \omega^{AB} \psi_B \right)  \bigg\} \,.
    \end{aligned}
\end{equation}
Also in this case $\cH = 0$, while the symplectic form and the first-class constraints keep the same form as in eqs.~\eqref{omega_5/2} and \eqref{F_5/2}.

As in section~\ref{sec:3/2}, the conjugate momenta can be eliminated enforcing the second-class constraints and one obtains the action
\be \label{S_E_5/2}
S_E = \int dt\, d^d x \left\{ \frac{1}{2}\left( \psi^\dag_A \omega^{AB} \dot{\psi}_B - \dot{\psi}^\dag_A \omega^{AB} \psi_B  \right) - \psi^\dag_{0k} \mathcal{F}^k- \mathcal{F}^{\dag k} \psi_{0k} \right\} ,
\ee
that can also be obtained directly from the action \eqref{spin52action} by taking the $c \to 0$ limit with the rescalings $\Psi_A = c^{1/2} \psi_A$ (i.e., $\Psi_{kl} = c^{1/2}\, \psi_{kl}$, $\Xi =  c^{1/2}\, \xi$) and $ \Psi_{0k} = c^{-1/2}\, \psi_{0k}$. This action is invariant under the gauge transformations
\be
    \delta \psi_{kl} =\partial_k \epsilon_l + \partial_l \epsilon_k \, , \qquad 
    \delta \xi = -2\, \gamma^k \gamma^l \partial_k \epsilon_l \, , \qquad
    \delta \psi_{0k} = \partial_t \epsilon_k \, ,
    \label{gauge_electric_5/2}
\ee
that can also be obtained from \eqref{gaugetransfospin5/2} by rescaling the gauge parameter as $\epsilon_k \to c^{1/2} \epsilon_k$.

\paragraph{Magnetic limit.}

As for Rarita–Schwinger fields, we wish to define a magnetic limit in which the canonical variables retain gauge transformations of the form \eqref{gaugetransfospin5/2}, possibly with independent gauge parameters (since the canonical pairs are no longer related by second-class constraints). To this end, the Lagrange multipliers can be redefined as
\begin{equation}
    \Psi_{0k} = \chi_k + \frac{1}{c}\, \zeta_k \, , \qquad 
    \lambda_{kl} = \tilde{\lambda}_{kl} - 4\, \partial_{(k} \zeta_{l)} \, , \qquad
    \lambda_\bullet = \tilde{\lambda}_\bullet + 4\, \gamma^k \gamma^l \partial_k \zeta_l \, .
\end{equation}
Note that the shift of the Lagrange multipliers can also be collectively rewritten as
\be
\lambda_A = \tilde{\lambda}_A - 2\, \delta_{\zeta} \Psi_A \, ,
\ee
where $\delta_\zeta \Psi_A$ takes the same form as a gauge transformation \eqref{gaugetransfospin5/2}. With these field redefinitions, the action becomes
\begin{equation}
\begin{split}
   S = \int dt\, d^d x & \bigg\{ \left( \Pi^A\dot{\Psi}_A + \dot{\Psi}^\dag_A \Pi^{\dag A}\right)  
    - \chi^\dag_k\mathcal{F}^k - \zeta^\dag_k \mathcal{G}^k - \mathcal{F}^{\dag k} \chi_k  - 
    \mathcal{G}^{\dag k} \zeta_k -\mathcal{H} \\
    & + \left( \Pi^A - \frac{1}{2c}\, \Psi^\dag_B \omega^{BA} \right) \tilde{\lambda}_A + \tilde{\lambda}_A^\dag\left( \Pi^{\dag A} + \frac{1}{2c}\, \omega^{AB} \Psi_B \right)  \bigg\} \,, 
\end{split}
\end{equation}
with
\begin{equation} \label{G_5/2}
   \mathcal{G}_k= - 4\, \partial_l \Pi^\dag{}_{\!k}{}^l + 4\, \gamma_k \gamma^l \partial_l \Theta^\dag \,, 
\end{equation}
while $\mathcal{H}$ and $\cF_k$ are the same as in eqs.~\eqref{H_5/2} and \eqref{F_5/2}. This rewriting follows from the identity
\be \label{relation_constraints_5/2}
4\, \partial_l \mathcal{C}^{\dag kl} - 4\, \gamma^k \gamma^l \partial_l \mathcal{C}^{\dag \bullet} = c^{-1}\, \mathcal{F}_k - \mathcal{G}_k 
\ee
where we collectively denoted the second-class constraints as
\be
\mathcal{C}^A := \Pi^A - \frac{1}{2c}\, \Psi^\dag_B \omega^{BA} \approx 0 \, .
\ee
This identity generalises \eqref{relation_constraints_3/2} and manifests that no new independent constraints are imposed.

Rescaling the Lagrange multipliers enforcing the second-class constraints as $\tilde{\lambda}_A =c^{\,\varepsilon}  \hat{\lambda}_A$ with $\varepsilon > 1$, in the limit $c \to 0$ one obtains the magnetic action
\begin{equation} \label{magnaction_spin5/2}
   S_M = \int dt\,d^d x \left\{ \left( \pi^A\dot{\psi}_A + \dot{\psi}^\dag_A \pi^{\dag A}\right)  
    - \chi^\dag_k\mathcal{F}^k - \zeta^\dag_k \mathcal{G}^k - \mathcal{F}^{\dag k} \chi_k  - 
    \mathcal{G}^{\dag k} \zeta_k  - \mathcal{H} \right\}, 
\end{equation}
where the Hamiltonian $\mathcal{H}$ as well as the constraints $\cF$ and $\cG$ take the same form as in eqs.~\eqref{H_5/2}, \eqref{F_5/2} and \eqref{G_5/2} with the relabellings $\Psi_A \to \psi_A$ and $\Pi^A \to \pi^A$.
These constraints continue to be of first class as before the limit,
\begin{equation}
\big\{ \mathcal{F}_k, \mathcal{F}^\dag_l \big\} = 0 \, , \qquad
\big\{ \mathcal{F}_k, \mathcal{G}^\dag_l \big\} = 0 \, , \qquad
\big\{ \mathcal{G}_k, \mathcal{F}^\dag_l \big\} = 0 \, , \qquad
\big\{ \mathcal{G}_k, \mathcal{G}^\dag_l \big\} = 0 \, ,
\end{equation}
and no secondary constraints are present:
\begin{equation} \label{secondary_5/2}
\big\{ \mathcal{F}[\epsilon], \mathcal{H} \big\} = 0\, , \qquad
\big\{ \mathcal{G}[\epsilon], \mathcal{H} \big\} =  2 \, \mathcal{F}^{\dag l} \gamma^0 \gamma^m \partial_l \epsilon_m \approx 0 \, ,
\end{equation}
where $\mathcal{F}[\epsilon]$ and $\mathcal{G}[\epsilon]$ are smeared constraints and $\epsilon_k$ is the smearing parameter.
The Hamiltonian satisfies $\{ \cH(x) , \cH(x') \} = 0$ as in the spin-$3/2$ case, while eqs.~\eqref{secondary_5/2} state that its Poisson bracket with the first-class constraints still weakly vanishes even if now $\cH$ is not anymore gauge invariant up to a boundary term. These conditions are consistent with the Carroll invariance of the action also in this case, in the sense discussed in \cite{Henneaux:2021yzg}.

As in section~\ref{sec:3/2}, with this procedure we doubled the number of independent spinor fields by removing the second-class constraints, but each spinor kept the same amount of gauge symmetry as before the limit, thus effectively doubling also the number of independent gauge parameters. Indeed, the first-class constraints generate the gauge transformations
\begin{subequations}
\begin{alignat}{5}
    \delta \psi_{kl}
    &=
    2\,\partial_{(k}\epsilon_{l)} \, , \qquad & 
    \delta \pi^{kl}
    &=
    -\frac{i}{2}
    \Big[
        2\,\partial^{(k}\rho^{\dagger l)}
        -\delta^{kl}\partial_m\rho^{\dagger m}
        -2\,\partial_m\rho^{\dagger (k|}
             \gamma^m\gamma^{|l)}
    \\
    & & &\quad\,
        -2\,\partial^{(k|}\rho^{\dagger m}
             \gamma_m\gamma^{|l)}
        +\delta^{kl}
         \partial_m\rho_n^\dagger\gamma^n\gamma^m
    \Big] \, ,
    \nonumber
    \\[5pt]
    \delta \xi
    &=
    -2\,\gamma^k\gamma^l\partial_k\epsilon_l \, , \qquad & 
    \delta \theta
    &=
    \frac{i}{2}
    \left[
        \partial_k\rho^{\dagger k}
        +
        \partial_l\rho^{\dagger k}\gamma_k\gamma^l
    \right] . 
\end{alignat}
\end{subequations}
which leave the action \eqref{magnaction_spin5/2} invariant when combined with the following transformations for the Lagrange multipliers:
\begin{equation}
    \delta \chi_k = \frac{1}{2}\,\partial_t \rho_k + \gamma^0 \gamma^l
        \partial_k \epsilon_l \, , \qquad 
    \delta \zeta_k = \frac{1}{2}\,\partial_t \epsilon_k \, .
\end{equation}

\paragraph{Minimal magnetic equations of motion.}

The equations of motion of the spin-$5/2$ Fang--Fronsdal action \eqref{spin52action} take the form
\begin{equation}
\begin{aligned}
 0 & =\gamma^\rho \partial_\rho \Psi^{\mu \nu}-\partial^\mu \slashed{\Psi}^\nu-\partial^\nu \slashed{\Psi}^\mu-\gamma^\mu \partial_\rho \Psi^{\rho \nu}-\gamma^\nu \partial_\rho \Psi^{\rho \mu} \\
& \quad +\gamma^\nu \gamma^\lambda \partial_\lambda \slashed{\Psi}^\mu+\gamma^\mu \gamma^\lambda \partial_\lambda \slashed{\Psi}^\nu+\eta^{\mu \nu} \partial^\sigma \slashed{\Psi}_\sigma \\
& \quad +\frac{1}{2}\left(\gamma^\nu \partial^\mu \Psi' +\gamma^\mu \partial^\nu \Psi' -\eta^{\mu \nu} \gamma^\rho \partial_\rho \Psi' \right),
\end{aligned}
\end{equation}
where we introduced the shorthand $\slashed{\Psi}_\mu := \gamma^\lambda \Psi_{\lambda\mu}$.
They can be used to define a magnetic limit in the same way as in the spin-$1/2$ and spin-$3/2$ cases, by acting on them with the projectors \eqref{projectors} and appropriately rescaling the field components before taking the limit $c \to 0$. To this end, it is convenient to implement the substitution
\begin{equation}
    \Psi_{00} = \Xi + 2\, \gamma^0 \gamma^k \Psi_{0k} \, , \label{redefspin5/2}
\end{equation}
suggested by the rewriting of the action in Hamiltonian form (cf.~\eqref{def_xi_5/2}). The decomposition of these equations into temporal and spatial components then proceeds as follows. The $(00)$ component gives
\begin{equation} \label{comp_00}
\begin{split}
 0 & = \frac{1}{2c} \left( \gamma^0 \partial_t \Xi -  \gamma^0 \partial_t \Psi_k{}^k \right) + \gamma^0 \left( \partial^k \Psi_{0 k} + \gamma^k  \gamma^l \partial_k \Psi_{0l} \right) \\
 & \quad -\frac{3}{2}\, \gamma^l \partial_l \Xi -\gamma^k \partial^l \Psi_{k l} +\frac{1}{2}\, \gamma^k \partial_k \Psi_{l}{}^l \,, 
\end{split}
\end{equation}
where we grouped the terms with time derivatives and those with an even or odd number of spatial gamma matrices, that behave differently under the action of the projectors \eqref{projectors}. In particular, the terms in the second line correspond to $-i \gamma^0\frac{\delta \cH}{\delta \Xi^\dagger}$. The components $(0k)$ give
\begin{equation} \label{comp_0k}
\begin{split}
 0 & = \frac{1}{2c} \left( \gamma_k \partial_t \Xi
- \gamma_k \partial_t \Psi^l_l \right)
+\gamma_k \partial_l \Psi_0^l +\gamma_k \gamma^l \gamma^m \partial_l \Psi_{0m} \\
& \quad+ \frac{1}{2}\, \gamma^0 \left( \partial_k \Xi
-2\, \partial_l \Psi_k{}^l
+ \partial_k \Psi_l{}^l + 2\, \gamma^l \gamma^m \partial_l \Psi_{km}
- 2\, \gamma_k \gamma^l \partial_l \Xi \right) , 
\end{split}
\end{equation}
while the components $(kl)$ give
\begin{align} 
0 & = 
\frac{1}{c}\, \Big( \gamma^0 \partial_t \Psi_{kl}
-\frac{1}{2}\, \delta_{kl} \gamma^0 \partial_t \Psi^{m}_{m}
+ 2\, \gamma_{(k|} \gamma^0 \gamma^m \partial_t \Psi_{|l)m} - \frac{1}{2}\, \delta_{kl} \gamma^0 \partial_t \Xi \Big)  -  i \gamma^0 \frac{\delta \mathcal{H}}{\delta \Psi^{\dag kl}} \label{comp_kl} \\
& \quad + \gamma^0 \Big( 2\, \gamma_{(k|} \gamma^m \partial_m \Psi_{|l)0}
+ \delta_{kl} \partial^m \Psi_{m0}
- 2\,\partial_{(k} \Psi_{l)0} + 2\, \gamma_{(k} \gamma^m \partial_{l)} \Psi_{0m}
- \delta_{kl} \gamma^m \gamma^n \partial_m \Psi_{0n} \Big) \, , \nonumber
\end{align}
with
\begin{equation}
\begin{split}
& i \gamma^0 \frac{\delta \mathcal{H}}{\delta \Psi^{\dag kl}} = -\,\gamma^m \partial_m \Psi_{kl}
+ 2\, \gamma^m \partial_{(k} \Psi_{l)m}
+ 2\, \gamma_{(k|} \partial_m \Psi_{|l)}{}^{m} -2\, \gamma_{(k|} \gamma^m \gamma^n \partial_m \Psi_{|l)n}
\\
& \qquad -\,\delta_{kl} \gamma^n \partial^m \Psi_{mn} - \gamma_{(k} \partial_{l)} \Psi_m{}^{m}
+\frac{1}{2}\, \delta_{kl} \gamma^m \partial_m \Psi_{n}{}^{n} + \gamma_{(k} \partial_{l)} \Xi
-\frac{1}{2}\, \delta_{kl} \gamma^m \partial_m \Xi \, .
\end{split}
\end{equation}
The constraints \eqref{F_5/2} can be recovered by contracting eq.~\eqref{comp_00} with $\gamma^k$, multiplying eq.~\eqref{comp_0k} by $\gamma^0$, and adding them together. In the following, we shall therefore manipulate eqs.~\eqref{comp_00} and \eqref{comp_kl}, and the constraints \eqref{F_5/2}.

Rescaling the field components as
\begin{equation}
     \Xi_\pm = c^{\mp 1/2} \xi_\pm \, , \qquad 
     \Psi_{kl \pm} =  c^{\mp 1/2} \psi_{kl \pm} \, , \qquad
     \Psi_{0i\pm} = c^{-1 \mp 1/2} \psi_{0i\pm} \, ,
\end{equation}
in the $c \rightarrow 0$ limit eq.~\eqref{comp_00} gives the triangular equations
\begin{subequations}
\begin{align}
 0 & = \frac{1}{2} \left( \gamma^0 \partial_t \xi_- -  \gamma^0 \partial_t \psi^k_{k -} \right) + \gamma^0 \left( \partial^k \psi_{0 k -} + \gamma^k  \gamma^l \partial_k \psi_{0l -} \right) - P_+ \left( i \gamma^0\frac{\delta \cH}{\delta \xi^\dagger} \right) , \label{first-traingular} \\[5pt]
 0 & = \frac{1}{2} \left( \gamma^0 \partial_t \xi_+ -  \gamma^0 \partial_t \psi^k_{k +} \right) + \gamma^0 \Big( \partial^k \psi_{0 k +} + \gamma^k  \gamma^l \partial_k \psi_{0l +} \Big) \, ,
\end{align}
\end{subequations}
where we stress that the last term in \eqref{first-traingular} depends on $\psi_{A +}$.
Eq.~\eqref{comp_kl} gives 
\begin{subequations}
\begin{align}
0 & = 
\Big( \gamma^0 \partial_t \psi_{kl-}
-\frac{1}{2}\, \delta_{kl} \gamma^0 \partial_t \psi^{m}_{m-}
+ 2\, \gamma_{(k|} \gamma^0 \gamma^m \partial_t \psi_{|l)m-} - \frac{1}{2}\, \delta_{kl} \gamma^0 \partial_t \xi_- \Big) - 2\,\gamma^0 \partial_{(k} \psi_{l)0-} \nonumber \\
& \quad + \gamma^0 \Big( 2\, \gamma_{(k|} \gamma^m \partial_m \psi_{|l)0-}
+ \delta_{kl} \partial^m \psi_{m0-}
+ 2\, \gamma_{(k} \gamma^m \partial_{l)} \psi_{0m-}
- \delta_{kl} \gamma^m \gamma^n \partial_m \psi_{0n-} \Big) \nonumber \\
& \quad - P_+ \Big(i \gamma^0 \, \frac{\delta \mathcal{H}}{\delta \psi^{\dag kl}} \Big) \, , \\[5pt]
0 & = 
\Big( \gamma^0 \partial_t \psi_{kl+}
-\frac{1}{2}\, \delta_{kl} \gamma^0 \partial_t \psi^{m}_{m+}
+ 2\, \gamma_{(k|} \gamma^0 \gamma^m \partial_t \psi_{|l)m+} - \frac{1}{2}\, \delta_{kl} \gamma^0 \partial_t \xi_+ \Big) - 2\,\gamma^0 \partial_{(k} \psi_{l)0+} \nonumber \\
& \quad + \gamma^0 \Big( 2\, \gamma_{(k|} \gamma^m \partial_m \psi_{|l)0+}
+ \delta_{kl} \partial^m \psi_{m0+}
+ 2\, \gamma_{(k} \gamma^m \partial_{l)} \psi_{0m+}
- \delta_{kl} \gamma^m \gamma^n \partial_m \psi_{0n+} \Big) \, ,
\end{align}
\end{subequations}
while the constraints keep all their terms when one projects them. They thus lead to the pair of equations 
\begin{equation}
    P_\pm \mathcal{F}_{k} = 0 \, .
\end{equation}

One can recover the same expressions from the equations of motion of the Carrollian action \eqref{magnaction_spin5/2} using the following truncation:
\begin{subequations}
\begin{alignat}{7}
P_+ \xi & = \xi_{+} \,, \quad & P_- \xi & = 0 \,, \quad &
P_+ \psi_{kl} & = \psi_{kl+} \,, \quad & P_- \psi_{kl} & = 0 \,, \\[5pt]
P_+ \theta^\dag & = 0 \,, \quad & P_- \theta^\dag & = - \omega^{\bullet | A} \psi_{A -} \,, \quad &
P_+ \pi^{\dag kl} & =0 \,, \quad & P_- \pi^{\dag kl} & = - \omega^{kl| A} \psi_{A -} \,, \\
P_- \zeta_{k} & = 0 \,, \quad & P_+ \zeta_{k} & = \frac{1}{2}\, \psi_{0k+} \,, \quad &
P_- \chi_{k} & = \psi_{0k -} \,, \quad & P_+ \chi_k & = 0 \,,
\end{alignat}
\end{subequations}
where we used the components of the symplectic two-form given in eq.~\eqref{omega_5/2}. Note that the same structures, such as the first-class constraints or the functional derivatives of the Hamiltonian, appear both in the projections of the Fang--Fronsdal equations and in the field equations of the magnetic action \eqref{magnaction_spin5/2}. In the next section, we shall take advantage of this rewriting to extend to fermions of arbitrary spin the matching between the minimal magnetic equations of motion and a truncation of those following from the magnetic action.

\section{Arbitrary spin} \label{sec:HS}

The Fang-Fronsdal action \cite{Fang:1978wz} for a massless field of spin $s+1/2$ is 
\begin{equation} \label{S_s}
    S = c^{-1} \int d^Dx \left\{ \frac{1}{2}\, \bPsi \slashed{\partial} \Psi + \frac{s}{2}\, \slashed{\bPsi}\slashed{\partial}\slashed{\Psi}  - \frac{1}{4} \binom{s}{2} \bPsi^\prime \slashed{\partial}\Psi^\prime + \binom{s}{2} \bPsi^\prime \partial \cdot \slashed{\Psi} - s\, \bPsi \partial \slashed{\Psi} \right\} + \text{h.c.} \, ,
\end{equation}
where the field is a complex-valued fully-symmetric spinor-tensor, $\Psi^\alpha{}_{\!\mu_1 \cdots \mu_s}= \Psi^\alpha{}_{\!(\mu_1 \cdots \mu_s)} $, whose indices are omitted here and in the following. Moreover, $\slashed{\Psi}:= \gamma^\rho \Psi_{\rho \mu_1 \cdots \mu_{s-1}}$ and $\Psi':= \eta^{\rho\sigma} \Psi_{\rho\sigma \mu_1 \cdots \mu_{s-2}}$. The shorthands $\partial \Psi$ and $\partial \cdot \Psi$ denote a symmetrised gradient and a divergence, where dividing by the number of terms in the symmetrisation is understood.
Starting from the spin-$7/2$ case, the field is bound to satisfy the algebraic constraint $\slashed{\Psi}^\prime = 0$. This action is invariant under the gauge transformations
\begin{equation} \label{gauge_s}
    \delta \Psi = s\, \partial \epsilon \,,
\end{equation}
where $\epsilon$ is a spinor-tensor with $s-1$ spacetime indices, satisfying the constraint $\slashed{\epsilon} = 0$. These variations thus generalise the gauge transformations of fields of spin $3/2$ and $5/2$ that we displayed, respectively, in eqs.~\eqref{gaugetransfospin3/2} and \eqref{gaugetransfospin5/2}. 

\paragraph{Action in Hamiltonian form.}

Also for arbitrary values of $s$, casting the action in Hamiltonian form essentially amounts to distinguishing the canonical variables from the Lagrange multipliers that enforce the first-class constraints generating the gauge symmetry.
As for spin-$5/2$ fields, one can identify the canonical variables as the field components whose gauge transformations do not contain any time derivative of the gauge parameter. This leads to identify them with the spatial components $\Psi_{k_1 \cdots k_s}$ and with the combination \cite{Campoleoni:2017vds} 
\begin{equation} \label{def-Xi_s}
    \Xi_{k_1 \cdots k_{s-2}} = \Psi_{00 k_1 \cdots k_{s-2}} - 2\, \gamma^0 \gamma^j \Psi_{0jk_1 \cdots k_{s-2}} \, .
\end{equation}
We then denote the Lagrange multipliers as $N_{k_1 \cdots k_{s-1}} := \Psi_{0 k_1 \cdots k_{s-1}}$ and from now on omitted indices are meant to be purely spatial (in contrast with the convention used only in eqs.~\eqref{S_s} and \eqref{gauge_s}). With this convention, the gauge transformations \eqref{gauge_s} read
\begin{subequations} \label{gauge_spatial}
\begin{align}
   & \delta \Psi = s\, \partial \epsilon \, , \\[5pt]
   & \delta \Xi = - 2\, \slashed{\partial}\slashed{\epsilon} - (s-2)\, \partial \epsilon^\prime \, , \\[5pt]
   & \delta N = c^{-1} \partial_t \epsilon + (s-1)\, \gamma^0 \partial \slashed{\epsilon} \, , 
\end{align}
\end{subequations}
where from now on $\slashed{\partial} := \gamma^k \partial_k$ and $\slashed{\epsilon}: = \gamma^l \epsilon_{lk_1 \cdots k_{s-2}}$. Moreover, as in \eqref{gaugetransfospin5/2}, we used the constraint $\gamma^\lambda \epsilon_{\lambda \mu_1 \cdots \mu_{s-2}} = 0$ to express everything in terms of $\epsilon_{k_1 \cdots k_{s-1}}$.
Using the algebraic constraint $\eta^{\mu\nu} \gamma^\rho \Psi_{\mu\nu\rho\sigma_1 \cdots \sigma_{s-3}} = 0$, all components of the field with more than two temporal components can be expressed in terms of the canonical variables and of the Lagrange multipliers as follows:
\begin{subequations}
\begin{align}
\Psi_{0 \cdots 0 k_{2 n+1} \cdots k_s} & = n\, \Xi_{k_{2 n+1} \cdots k_s}^{[n-1]}+ 2 n\, \gamma^0 \slashed{N}_{k_{2 n+1} \cdots k_s}^{[n-1]} - (n-1)\, \Psi_{k_{2 n+1} \cdots k_s}^{[n]} \, ,  \\[5pt]
\Psi_{0 \cdots 0 k_{2 n+2} \cdots k_s} & = n\, \gamma^0 \slashed{\Xi}_{k_{2 n+2} \cdots k_s}^{[n-1]} + (2n+1)\, N_{k_{2 n+2} \cdots k_s}^{[n]} - n\, \gamma^0 \slashed{\Psi}_{k_{2 n+2} \cdots k_s}^{[n]} \, ,
\end{align}
\end{subequations}
where $X^{[n]}$ denotes the $n$-th spatial trace of the corresponding field (with the convention that terms with a negative number of traces are absent). For instance, $\Psi^{[n]}:= \delta^{i_1j_1} \cdots \delta^{i_nj_n} \Psi_{i_1j_1 \cdots i_nj_n k_1 \cdots k_{s-2n}}$.
This allows one to rewrite the action in canonical form as
\be \label{S_H_s}
  S=\int dt\,d^d x \left\{ \left( \Pi^A \dot{\Psi}_A + \dot{\Psi}_A^{\dagger} \Pi^{\dag A}\right) - N^{\dagger} \mathcal{F} - \mathcal{F}^{\dagger} N - \mathcal{H} + \cC^A \lambda_A + \lambda^\dag_A \cC^{\dag A} \right\} \, ,
\ee
where we already introduced a canonical momentum for each canonical variable,
\begin{equation}
\Psi_A=\left(\begin{array}{c} \Psi_{k_1 \cdots k_s} \\ \Xi_{k_1 \cdots k_{s-2}} \end{array}\right) , \quad
\Pi^A := \frac{\delta \cL}{\delta \dot{\Psi}_A} = \left(\begin{array}{c} \Pi^{k_1 \cdots k_s} \\ \Theta^{k_1 \cdots k_{s-2}} \end{array}\right) , 
\end{equation}
as well as the explicit dependence on the second-class constraints. Using the index-free convention, the first-class constraints read
\begin{equation} \label{F_s}
\begin{split}
\mathcal{F} & = \frac{i}{2} \sum_{n=0}^{[s / 2]} \binom{s}{2 n} \Big\{2 n\, \gamma\, \delta^{n-1}  \Big[ \slashed{\partial} \Xi^{[n-1]} + (s-2 n)\, \partial \slashed{\Xi} ^{[n-1]} + 2(n-1)\, \partial \cdot \slashed{\Xi}^{[n-2]} \\
& - \slashed{\partial} \Psi^{[n]} + (s-2 n)\, \partial \slashed{\Psi}^{[n]}+2 n\, \partial \cdot \slashed{\Psi}^{[n-1]} \Big] + (s-2 n)\, \delta^n \Big[ 2 n\,\partial \cdot \Xi^{[n-1]} \\
& + (s-2 n-1)\, \partial \Xi^{[n]} + 2(n-1)\, \partial \cdot \Psi^{[n]} + (s-2 n-1)\, \partial \Psi^{[n+1]}+ 2\, \slashed{\partial} \slashed{\Psi}^{[n]}\Big]\Big\} \approx 0 \, ,
\end{split}
\end{equation}
where, e.g., $\gamma \Psi := \gamma_{(k_1} \Psi_{k_2 \cdots k_{s+1})}$, while the second-class constraints read
\be
\mathcal{C}^A := \Pi^A - \frac{1}{2c}\, \Psi^\dag_B \omega^{BA} \approx 0 \, ,
\ee
where $\omega^{AB}$ denotes the symplectic form, given in appendix~\ref{app:H} together with the Hamiltonian $\mathcal{H}$. When rewritten in the canonical form \eqref{S_H_s}, the action has the same structure as the actions \eqref{Hamiltonian form action 3/2} and \eqref{S_H_5/2}; all technical complications brought by the additional indices are hidden in the explicit expressions for the symplectic form, the first-class constraints and the Hamiltonian. Since the form of these objects is modified neither in the electric nor in the magnetic limit, the definition of these limits proceeds along the same lines as in the previous sections. 

\paragraph{Electric limit.} In this case, we rescale the dynamical variables in a way that leaves the kinetic term invariant while removing the relative factor of $c$ in the second-class constraints. Preserving all constraints (first and second class) then requires one to rescale appropriately the Lagrange multipliers. All in all, one has to rescale the fields as
\begin{equation}
    \Psi_A = \sqrt{c} \, \psi_A \,, \qquad 
    \Pi^A = \frac{1}{\sqrt{c}}\, \pi^A \,, \qquad
    N =  \frac{1}{\sqrt{c}}\, \tilde{N} \,, \qquad 
    \lambda_A = \sqrt{c}\, \tilde{\lambda}_A \,.
 \end{equation}
Sending $c \rightarrow 0$, one obtains the action
\be
\begin{split}
  S_E =\int dt\,d^d x & \bigg\{ \left( \pi^A \dot{\psi}_A + \dot{\psi}_A^{\dagger} \pi^{\dag A}\right) - \tilde{N}^{\dagger} \mathcal{F} - \mathcal{F}^{\dagger} \tilde{N} \\
  & + \left( \pi^A - \frac{1}{2}\, \psi^\dag_B \omega^{BA} \right) \tilde{\lambda}_A + \tilde{\lambda}_A^\dag\left( \pi^{\dag A} + \frac{1}{2}\, \omega^{AB} \psi_B \right)  \bigg\} \,.
\end{split}
\ee
Eliminating the conjugate momenta through the second-class constraints one obtains the action
\be
  S_E =\int dt\,d^d x \left\{ \frac{1}{2}\left( \psi^\dag_A \omega^{AB} \dot{\psi}_B - \dot{\psi}^\dag_A \omega^{AB} \psi_B  \right) - \tilde{N}^{\dagger} \mathcal{F} - \mathcal{F}^{\dagger} \tilde{N} \right\} ,
\ee
that can also be obtained directly from the action \eqref{S_s} by taking the $c \to 0$ limit with the rescalings $\Psi_A = c^{1/2} \psi_A$ and $N = c^{-1/2} \tilde{N}$. This action is invariant under the gauge transformations obtained from \eqref{gauge_spatial} by rescaling the gauge parameter as $\epsilon_{k_1 \cdots k_{s-1}} \to c^{1/2} \epsilon_{k_1 \cdots k_{s-1}}$.

\paragraph{Magnetic limit.} In this case,  we do not rescale the canonical variables, and we introduce the following redefinition of the Lagrange multipliers:
\begin{equation} \label{N-redef_s}
    N = \chi + \frac{1}{c}\, \zeta \, , \qquad 
    \lambda_A = \tilde{\lambda}_A - 2\, \delta_{\zeta} \Psi_A \, ,
\end{equation}
where $\delta_{\zeta} \Psi_A$ has the same form as a gauge transformation, with the gauge parameter substituted by the spinor-tensor $\zeta_{k_1 \cdots k_{s-1}}$.
With these field redefinitions, the action becomes
\begin{equation}
\begin{split}
   S = \int dt\, d^d x & \bigg\{ \left( \Pi^A\dot{\Psi}_A + \dot{\Psi}^\dag_A \Pi^{\dag A}\right)  
    - \chi^\dag \mathcal{F} - \zeta^\dag \mathcal{G} - \mathcal{F}^{\dag} \chi  - 
    \mathcal{G}^{\dag} \zeta -\mathcal{H} \\
    & + \left( \Pi^A - \frac{1}{2c}\, \Psi^\dag_B \omega^{BA} \right) \tilde{\lambda}_A + \tilde{\lambda}_A^\dag\left( \Pi^{\dag A} + \frac{1}{2c}\, \omega^{AB} \Psi_B \right)  \bigg\} \,, 
\end{split}
\end{equation}
with 
\begin{equation} \label{G_s}
  \mathcal{G} := 2 \left( - s\, \partial\cdot \Pi^{\dag} + 2\,  \gamma\, \slashed{\partial} \Theta^{\dag} + (s-2)\, \delta\, \partial\cdot \Theta^{\dag} \right) \approx 0 \, ,
\end{equation}
while $\cF$ and the Hamiltonian remain the same as in eqs.~\eqref{F_s} and \eqref{H_s}. The form of the field redefinition \eqref{N-redef_s} and of the constraints \eqref{G_s} can be derived by rewriting the gauge transformations of the canonical variables in operatorial form as follows:
\be \label{gauge_op}
\delta \Psi_A = \cD_A{}^{k_1 \cdots k_{s-1}} \epsilon_{k_1 \cdots k_{s-1}} \, .
\ee
The first-class constraints \eqref{F_s} can then be rewritten as
\be
\cF^{k_1 \cdots k_{s-1}} = - \big(\cD^*){}^{k_1 \cdots k_{s-1}}{}_{\!A} \, \omega^{AB} \Psi_B \, ,
\ee
where $\cD^*$ denotes the formal adjoint of the differential operator $\cD$, in which derivatives are ``integrated by parts'' and the order of gamma matrices is reversed. More precisely,
\be \label{def_Dstar}
\int d^d x \left( \cD_A{}^{k_1 \cdots k_{s-1}} \zeta_{k_1 \cdots k_{s-1}} \right) ^\dag X^A = \int d^d x \, \zeta_{k_1 \cdots k_{s-1}}^\dag \big(\cD^*\big){}^{k_1 \cdots k_{s-1}}{}_A X^A \, .
\ee
The counterpart \eqref{G_s} of the first-class constraints reads
\be
\cG^{k_1 \cdots k_{s-1}} = 2 \big(\cD^*\big){}_A{}^{k_1 \cdots k_{s-1}} \Pi^{\dag A} \, .
\ee
These identities allow one to generalise \eqref{relation_constraints_5/2} to 
\be \label{relation_constraints_s}
-2 \big(\cD^*\big){}_A{}^{k_1 \cdots k_{s-1}} \cC^{\dag A} = c^{-1} \cF^{k_1 \cdots k_{s-1}} - \cG^{k_1 \cdots k_{s-1}}  \, .
\ee

Rescaling the Lagrange multipliers enforcing the second-class constraints as $\tilde{\lambda}_A =c^{\,\varepsilon}  \hat{\lambda}_A$ with $\varepsilon > 1$, in the limit $c \to 0$ one then obtains the magnetic action
\begin{equation} \label{magnaction_spin-s}
   S_M = \int dt\,d^d x \left\{ \left( \pi^A\dot{\psi}_A + \dot{\psi}^\dag_A \pi^{\dag A}\right)  
    - \chi^\dag\mathcal{F} - \zeta^\dag \mathcal{G} - \mathcal{F}^{\dag} \chi  - 
    \mathcal{G}^{\dag} \zeta  - \mathcal{H} \right\}, 
\end{equation}
where the Hamiltonian and all constraints take the same form as in the relativistic theory with the relabellings $\Psi_A \to \psi_A$ and $\Pi^A \to \pi^A$. This action is invariant under the gauge transformations
\be
\delta \psi_A = \cD_A{}^{k_1 \cdots k_{s-1}} \epsilon_{k_1 \cdots k_{s-1}} \, , \qquad
\delta \pi^{\dag A} = - \frac{1}{2}\, \omega^{AB} \cD_B^{k_1 \cdots k_{s-1}} \rho_{k_1 \cdots k_{s-1}} \, ,
\ee
generated by the first-class constraints, together with the following transformations for the Lagrange multipliers:
\begin{subequations}
\begin{align}
    \delta \chi_{k_1 \cdots k_{s-1}} & = \frac{1}{2}\,\partial_t \rho_{k_1 \cdots k_{s-1}} + (s-1)\,\gamma^0 \gamma^l
        \partial_{(k_1} \epsilon_{k_2 \cdots k_{s-1})l} \, , \\ 
    \delta \zeta_{k_1 \cdots k_{s-1}} & = \frac{1}{2}\,\partial_t \epsilon_{k_1 \cdots k_{s-1}} \, .
\end{align}
\end{subequations}

\paragraph{Minimal magnetic equations of motion.}

The differential operator that we introduced in eq.~\eqref{gauge_op} allows one to define conveniently the minimal magnetic truncation of the Fang--Fronsdal equations and to compare it with the equations of motion following from the magnetic action \eqref{magnaction_spin-s}. Indeed, the rewriting \eqref{Fang-Fronsdal_H} of the Fang--Fronsdal action \eqref{S_s} shows that its equations of motion can be cast in the form
\begin{subequations}
\begin{align}
& c^{-1}\, \omega^{AB} \partial_t \Psi_B - \omega^{AB} \cD_B{}^{k_1 \cdots k_{s-1}} N_{k_1 \cdots k_{s-1}} - H^{AB} \Psi_B = 0 \, , \\[5pt]
& \big(\cD^*\big){}^{k_1 \cdots k_{s-1}}{}_{\!A} \omega^{AB} \Psi_B = 0 \, ,
\end{align}
\end{subequations}
where we introduced the self-adjoint\footnote{Strictly speaking, this operator is self-adjoint only if one also allows for integrations by parts of the derivatives in the sense of eq.~\eqref{def_Dstar}.} operator $H^{AB}$ via
\be
\cH = \Psi_A^\dag H^{AB} \Psi_B \, .
\ee
As displayed in appendix~\ref{app:H}, all addenda in the symplectic form $\omega^{AB}$ and in the operator $\cD_A^{k_1 \cdots k_{s-1}}$ contain either zero or two spatial gamma matrices, while the terms in the operator $H^{AB}$ contain either one or three spatial gamma matrices. As a result, acting with the projectors \eqref{projectors} and rescaling the fields as
\be
\Psi_\pm = c^{\mp \frac{1}{2}} \psi_\pm \, , \qquad
N_{k_1 \cdots k_{s-1} \pm} = c^{-1\mp\frac{1}{2}} \tilde{N}_{k_1 \cdots k_{s-1} \pm} \, ,
\ee
one obtains the equations
\begin{subequations}
\begin{align}
& \omega^{AB} \left( \partial_t \psi_{B +} - \cD_B{}^{k_1 \cdots k_{s-1}} \tilde{N}_{k_1 \cdots k_{s-1} +} \right) = 0 \, , \label{minimal-magn_1} \\
& \omega^{AB} \left( \partial_t \psi_{B -} -  \cD_B{}^{k_1 \cdots k_{s-1}} \tilde{N}_{k_1 \cdots k_{s-1} -} \right) - H^{AB} \psi_{B +} = 0 \, , \label{minimal-magn_2} \\[2pt]
& \big(\cD^*\big){}^{k_1 \cdots k_{s-1}}{}_{\!A} \omega^{AB} \psi_{B \pm} = 0 \, . \label{minimal-magn_3}
\end{align}
\end{subequations}

The equations of motion of the magnetic action \eqref{magnaction_spin-s} are instead
\begin{subequations}
\begin{align}
& \partial_t \psi_A - 2\,\cD_A{}^{k_1 \cdots k_{s-1}} \zeta_{k_1 \cdots k_{s-1}} = 0 \, , \label{magn_1} \\[5pt]
& - \partial_t \pi^{\dag A} - \omega^{AB} \cD_B{}^{k_1 \cdots k_{s-1}} \chi_{k_1 \cdots k_{s-1}} - H^{AB} \psi_B = 0 \, , \label{magn_2} \\[5pt]
& \big(\cD^*\big){}^{k_1 \cdots k_{s-1}}{}_{\!A} \omega^{AB} \psi_B = 0 \, , \label{magn_3} \\[5pt]
& \big(\cD^*\big){}^{k_1 \cdots k_{s-1}}{}_{\!A} \pi^{\dag A} = 0 \, . \label{magn_4}
\end{align}
\end{subequations}
Imposing the truncation
\begin{subequations}
\begin{alignat}{5}
P_+ \psi_A & = \psi_{A+} \, , & \qquad P_- \psi_A & = 0 \, , & \qquad 
P_+ \pi^{\dag A} & = 0 \, , & \qquad P_- \pi^{\dag A} & = - \omega^{AB} \psi_{B -}  \, , \\
P_+ \chi & = 0 \, , & \qquad P_- \chi & = \tilde{N}_- \, , & \qquad 
P_+ \zeta & = \frac{1}{2}\,\tilde{N}_{+} \, , & \qquad P_- \zeta & = 0 \, ,
\end{alignat}
\end{subequations}
the $P_+$ projection of eq.~\eqref{magn_1} gives eq.~\eqref{minimal-magn_1}, the $P_-$ projection of eq.~\eqref{magn_2} gives eq.~\eqref{minimal-magn_2}, while the constraint \eqref{magn_3} gives the $P_+$ projection of \eqref{minimal-magn_3} and \eqref{magn_4} gives its $P_-$ projection.

\section{Conclusions} \label{sec:conclusions}

In this paper, we have defined electric and magnetic Carrollian free field theories for fermions of arbitrary spin in any spacetime dimension $D \geq 3$. These theories arise as inequivalent $c\to0$ limits of the Lorentz-invariant Fang–Fronsdal actions. The electric limit is obtained through a rescaling of the field components that preserves both the kinetic term and the first-class constraints generating the gauge symmetry of the Fang–Fronsdal action, at the price of a vanishing Hamiltonian in the limit. In the magnetic case, we first cast the action in Hamiltonian form and then defined the limit so as to remove the second-class constraints of the relativistic theory. As in the magnetic limit of bosonic fields, the conjugate momenta can therefore no longer be eliminated algebraically, effectively doubling the number of independent spinor fields. Nevertheless, we have shown that the gauge transformations of all canonical variables can retain the same form as their relativistic counterparts. The equations of motion derived from the magnetic action form a reducible but indecomposable system. We have also shown that this system can be consistently truncated to a smaller one with similar properties, which can be obtained directly by taking a suitable magnetic limit of the Fang–Fronsdal equations of motion.

The treatment of the first-class constraints in the magnetic case admits a certain freedom, owing to the doubling of the independent spinor fields. Although our proposal has the advantage of assigning a gauge symmetry to all spinor fields, a natural next step would be to test it by investigating the supersymmetry transformations that could relate our actions to the magnetic Carrollian bosonic actions of \cite{Henneaux:2021yzg}. This could be done, for instance, by studying the Carrollian limits of the higher-spin supermultiplets of \cite{Curtright:1979uz, Kuzenko:1993jp, Kuzenko:1993jq}, in the spirit of the analysis of the Wess–Zumino multiplet presented in \cite{Koutrolikos:2023evq}.
Carrollian spin-$1/2$ fermions have already been coupled to electric and magnetic Carroll gravity (see, e.g., \cite{Bergshoeff:2017btm, Bergshoeff:2023vfd}), while the electric and magnetic Carroll limits of $D=4$, $\cN=1$ supergravity have recently been analysed in \cite{Henneaux:2026dfc}. It would be worthwhile to determine whether Carrollian supergravity theories could be constructed from scratch starting from the magnetic spin-$3/2$ model introduced here, and whether this construction would enlarge the class of models presented in \cite{Henneaux:2026dfc}.
Finally, it would be interesting to extend the analysis of Carrollian limits to massive bosonic and fermionic field theories of arbitrary spin, starting, for instance, from the actions of \cite{Singh:1974qz, Singh:1974rc}, which provide the natural massive counterparts of the massless actions employed here and in \cite{Henneaux:2021yzg}.

\acknowledgments

We thank Nicolas Boulanger, Daniel Grumiller, Marc Henneaux, Nicolas Maindiaux, Luciano Montecchio, Sylvain Thom\'ee and Matthieu Vilatte for discussions. This work was supported by the Fonds de la Recherche Scientifique -- FNRS under Grants nr.\ T.0047.24 and FC.55077. The work of L.M.\ was partially supported by a Junior Research Fellowship of the Erwin Schr\"odinger International Institute for Mathematics and Physics and she thanks TU Wien for hospitality. 

\appendix

\section{Conventions} \label{App:Conventions}

We take the $\gamma$-matrices to obey the Clifford algebra
\begin{align}
    \{\gamma_{\mu}, \gamma_{\nu}\} = 2\, \eta_{\mu\nu} \mathds{1} \,,
\end{align}
where $\eta_{\mu\nu}$ is the $D$-dimensional Minkowski metric with mostly plus signature.
They satisfy the Hermiticity properties
\begin{align}
    \big( \gamma^0 \big)^{\dagger} = - \gamma^0 \, , \qquad
    \big( \gamma^k \big)^{\dagger} = \gamma^k \,, \qquad  (k = 1,2,\cdots, d = D-1) \,.
\end{align}

The Dirac conjugate of a spinor is defined as
\begin{eqnarray}
    \bar{\Psi} := i\, \Psi^{\dagger} \gamma^0 \,.
\end{eqnarray}
The complex conjugate of a product of (Grassmann-valued) spinor components $\varepsilon_\alpha$, $\psi_\beta$ is defined as the product of their complex conjugates in reverse order:
\begin{eqnarray}
    (\varepsilon_{\alpha} \psi_{\beta})^{*} =  \psi_{\beta}^{*} \varepsilon_{\alpha}^{*} \, . 
\end{eqnarray}

We note the following useful commutation properties of the projectors $P_{\pm}$ introduced in \eqref{projectors}:
\begin{equation}
    P_{\pm} \gamma^0 = \gamma^0 P_{\pm} \, , \qquad
    P_{\pm} \gamma^a = \gamma^a P_{\mp} \, .
\end{equation}
Furthermore, projected spinors obey the following properties:
\begin{eqnarray}
    \bar{\Psi}_{\pm} = \bar{\Psi}_{\pm} P_{\pm} \,, \qquad \qquad \qquad \gamma^0 \Psi_{\pm} = \mp\, i\, \Psi_{\pm} \,.
\end{eqnarray}

Indices enclosed between a pair of round (square) brackets are meant to be (anti)symmetrised and dividing by the number of terms in the sum is understood. For instance,
\be
\gamma^{\mu\nu} = \gamma^{\left[\mu\right.} \gamma^{\left.\nu\right]} = 
\frac{1}{2} \left( \gamma^{\mu} \gamma^\nu - \gamma^{\nu} \gamma^{\mu} \right) .
\ee

\section{Fang--Fronsdal action in Hamiltonian form} \label{app:H}

The Fang--Fronsdal action \eqref{S_s} can be cast in the Hamiltonian form
\be \label{Fang-Fronsdal_H}
S = \int dt\, d^d x \left\{ \frac{1}{2c}\left( \Psi^\dag_A \omega^{AB} \dot{\Psi}_B - \dot{\Psi}^\dag_A \omega^{AB} \Psi_B  \right) - N_{k(s-1)}^\dag \mathcal{F}^{k(s-1)} - \mathcal{F}^{\dag k(s-1)} N_{k(s-1)} - \mathcal{H} \right\} .
\ee
where a label followed by a number between round brackets denotes a set of symmetrised indices, e.g., $N_{k(s-1)} := N_{k_1 \cdots k_{s-1}}$. Moreover, repeated covariant or contravariant indices denote a symmetrisation, e.g., $A_k B_{k(2)} := A_{(k_1} B_{k_2k_3)}$. For simplicity, we never used this notation in the main body of the text, but it is convenient to display explicitly the symplectic form $\omega^{AB}$ and the Hamiltonian $\cH$. With these conventions
\be
\Psi_A= \left(\begin{array}{c} \Psi_{k(s)} \\ \Xi_{k(s-2)} \end{array}\right) ,  \qquad
\omega^{A B}=\left(\begin{array}{cc}
\omega^{k(s) \mid l(s)} & \omega^{k(s) \mid \bullet i(s-2)} \\
\omega^{\bullet j(s-2) \mid l(s)} & \omega^{\bullet j(s-2) \mid \bullet i(s-2)}
\end{array}\right) ,
\ee
where $\Xi_{k(s-2)} := \Psi_{00 k(s-2)} - 2\, \gamma^0 \gamma^j \Psi_{0jk(s-2)}$ as in eq.~\eqref{def-Xi_s}.
The components of the symplectic form, which satisfies the property $\omega^{AB} = - \left(\omega^{BA}\right)^\dag$, are given by 
\begingroup
\allowdisplaybreaks
\begin{align} 
\omega^{k(s) \mid l(s)}
&=
\sum_{n=0}^{\lfloor s/2 \rfloor}
\binom{s}{2n}
\Bigg\{
i\left(
\frac{\frac{3}{2}n+1}{2n+1}
\right)(s-2n)\,
\gamma^{k}\gamma^{l}
\big(\eta^{kl}\big)^{s-2n-1}
\big(\eta^{kk}\big)^n
\big(\eta^{ll}\big)^n
\nonumber\\*
&\hspace{2.5cm}
-i\left(-\frac{3}{2}n+1\right)
\big(\eta^{kk}\big)^n
\big(\eta^{ll}\big)^n
\big(\eta^{kl}\big)^{s-2n}
\Bigg\} \, ,
\\[1em]
\omega^{k(s) \mid \bullet\,i(s-2)}
&=
\sum_{n=0}^{\lfloor s/2 \rfloor}
\binom{s}{2n}
\Bigg\{
\frac{i}{2}\,
\frac{n(s-2n)}{2n+1}\,
\gamma^{k}\gamma^{i}
\big(\eta^{kk}\big)^n
\big(\eta^{ii}\big)^{n-1}
\big(\eta^{ki}\big)^{s-2n-1}
\nonumber\\*
&\hspace{2.5cm}
+\frac{i}{2}\,n
\big(\eta^{ki}\big)^{s-2n}
\big(\eta^{kk}\big)^n
\big(\eta^{ii}\big)^{n-1}
\Bigg\} \, ,
\\[1em]
\omega^{\bullet\,j(s-2) \mid l(s)}
&=
\sum_{n=0}^{\lfloor s/2 \rfloor}
\binom{s}{2n}
\Bigg\{
\frac{i}{2}\,
\frac{n(s-2n)}{2n+1}\,
\big(\eta^{jj}\big)^{n-1}
\big(\eta^{ll}\big)^n
\big(\eta^{jl}\big)^{s-2n-1}
\gamma^{j}\gamma^{l}
\nonumber\\*
&\hspace{2.5cm}
+\frac{i}{2}\,n
\big(\eta^{jj}\big)^{n-1}
\big(\eta^{jl}\big)^{s-2n}
\big(\eta^{ll}\big)^n
\Bigg\} \, ,
\\[1em]
\omega^{\bullet\,j(s-2) \mid \bullet\,i(s-2)}
&=
\sum_{n=0}^{\lfloor s/2 \rfloor}
\binom{s}{2n}
\Bigg\{
-\frac{i}{2}\,
\frac{n(s-2n)}{2n+1}\,
\big(\eta^{ij}\big)^{s-2n-1}
\big(\eta^{jj}\big)^{n-1}
\big(\eta^{ii}\big)^{n-1}
\gamma^{j}\gamma^{i}
\nonumber\\*
&
-\frac{i}{4}\,
\frac{(s-2n)(s-2n-1)}{2n+1}\,
\big(\eta^{ij}\big)^{s-2n-2}
\big(\eta^{jj}\big)^n
\big(\eta^{ii}\big)^n
\Bigg\} \, .
\end{align}
\endgroup
In the current notation, the first-class constraints \eqref{F_s} read
\begin{equation} \label{F_s_app}
\begin{split}
\mathcal{F} & = \frac{i}{2} \sum_{n=0}^{[s / 2]} \binom{s}{2 n} \Big\{2 n\, \gamma_k (\delta_{kk})^{n-1}  \Big[ \slashed{\partial} \Xi^{[n-1]}_{k(s-2n)} + (s-2 n)\, \partial_k \slashed{\Xi}^{[n-1]}_{k(s-2n-1)} \\
& + 2(n-1)\, \partial \cdot \slashed{\Xi}^{[n-2]}_{k(s-2n)} - \slashed{\partial} \Psi^{[n]}_{k(s-2n)} + (s-2 n)\, \partial_k \slashed{\Psi}^{[n]}_{k(s-2n-1)} +2 n\, \partial \cdot \slashed{\Psi}^{[n-1]}_{k(s-2n)} \Big] \\
& + (s-2 n)\, (\delta_{kk})^n \Big[ 2 n\,\partial \cdot \Xi^{[n-1]}_{k(s-2n-1)} + (s-2 n-1)\, \partial_k \Xi^{[n]}_{k(s-2n-2)} \\
& + 2(n-1)\, \partial \cdot \Psi^{[n]}_{k(s-2n-1)} + (s-2 n-1)\, \partial_k \Psi^{[n+1]}_{k(s-2n-2)} + 2\, \slashed{\partial} \slashed{\Psi}^{[n]}_{k(s-2n-1)}\Big]\Big\} \approx 0 \, ,
\end{split}
\end{equation}
where we stress that the slash symbol denotes here a contraction with a spatial gamma matrix $\gamma^k$, and that a dot denotes a contraction over spatial indices as, e.g., in eq.~\eqref{gauge_spatial}.
Omitting for simplicity the $n$-dependent groups of contracted indices (so that, for instance, $\bar{\Xi}^{[n-1]}_{j\,k(s-2n-1)}
\gamma^{j}\slashed{\partial}
\slashed{\Xi}^{[n-1]\,k(s-2n-1)} \to \bar{\Xi}^{[n-1]}_{j}
\gamma^{j}\slashed{\partial}
\slashed{\Xi}^{[n-1]}$), the Hamiltonian reads
%
\begin{equation}
\begin{split} 
\mathcal{H}
& = -\sum_{n=0}^{\lfloor s/2 \rfloor}
\binom{s}{2n}
\Bigg\{
-\frac{3n}{2}\,
\bar{\Xi}^{[n-1]}
\slashed{\partial}
\Xi^{[n-1]}
-\frac{(s-2n)n^{2}}{2n+1}\,
\bar{\Xi}^{[n-1]}_{j}
\gamma^{j}\slashed{\partial}
\slashed{\Xi}^{[n-1]}
\\
&{}+\frac{1}{4}(s-2n)2n\,
\bar{\Xi}^{[n-1]}_{j}
\gamma^{j}\slashed{\partial}
\slashed{\Xi}^{[n-1]}
-(s-2n)2n\,
\bar{\Xi}^{[n-1]}_{j}
\gamma^{j}\partial^{i}
\Xi^{[n-1]}_{i}
\\
&{}+2n(2n-2)\,
\bar{\Xi}^{[n-2]}_{jl}
\gamma^{j}\partial^{l}
\Xi^{[n-1]}
-n(n-1)\,
\bar{\Xi}^{[n-1]}
\slashed{\partial}
\Psi^{[n]}
\\
&\hspace{0pt}
-n(n-1)\,
\bar{\Psi}^{[n]}
\slashed{\partial}
\Xi^{[n-1]}
{}+\frac{1}{4}2n(2n-1)(s-2n+1)\,
\bar{\Xi}^{[n-1]}
\slashed{\partial}
\Psi^{[n]}
\\
&\hspace{0pt}
+\frac{1}{4}2n(2n-1)(s-2n+1)\,
\bar{\Psi}^{[n]}
\slashed{\partial}
\Xi^{[n-1]}
-\frac{1}{2}(s-1-2n)(s-2n)\,
\bar{\Xi}^{[n]}
\partial^{i}
\slashed{\Psi}^{[n]}_{i} \\
 &-(s-2n)n\,
{ \bar{\Xi}^{[n-1]}_{l}
\partial^{l}
\slashed{\Psi}^{[n]}} -\frac{1}{2}(s-2n-1)(s-2n)\,
\bar{\Psi}^{[n]}_{ij}
\gamma^{j}\partial^{i}
\Xi^{[n]} \\
&-(s-2n)n\,
\bar{\Psi}^{[n]}_{i}
\gamma^{i}\partial^{l}
\Xi^{[n-1]}_{l} 
+\frac{(s-2n)n^{2}}{2n+1}
\left[
\bar{\Xi}^{[n-1]}_{j}
\gamma^{j}\slashed{\partial}
\slashed{\Psi}^{[n]}
+\bar{\Psi}^{[n]}_{j}
\gamma^{j}\slashed{\partial}
\slashed{\Xi}^{[n-1]}
\right]
\\
&{}-\frac{1}{4}2n(s-2n)
\left[
\bar{\Xi}^{[n-1]}_{j}
\gamma^{j}\slashed{\partial}
\slashed{\Psi}^{[n]}
+\bar{\Psi}^{[n]}_{j}
\gamma^{j}\slashed{\partial}
\slashed{\Xi}^{[n-1]}
\right]
\\
&{}+\frac{1}{2}2n(s-2n)
\left[
\bar{\Psi}^{[n]}_{j}
\gamma^{j}\partial^{i}
\Xi^{[n-1]}_{i}
+\bar{\Xi}^{[n-1]}_{i}
\gamma^{j}\partial^{i}
\Psi^{[n]}_{j}
\right]
\\
&{}+\frac{n(s-1-2n)}{2n+1}
\left[
\bar{\Psi}^{[n]}_{jl}
\gamma^{j}\partial^{l}
\Xi^{[n]}
+\bar{\Xi}^{[n]}
\gamma^{j}\partial^{l}
\Psi^{[n]}_{jl}
\right]
\\
&{}+\left(-\frac{3n}{2}+1\right)
\bar{\Psi}^{[n]}
\slashed{\partial}
\Psi^{[n]}
+(s-2n)\frac{5n+2}{4n+2}\,
\bar{\Psi}^{[n]}_{l}
\gamma^{l}\slashed{\partial}
\slashed{\Psi}^{[n]}
\\
&{}+\frac{1}{2}(s-1-2n)(s-2n)\,
{\bar{\Psi}^{[n+1]}
\partial^{i}
\slashed{\Psi}^{[n]}_{i}}
+(s-2n)(n-1)\,
\bar{\Psi}^{[n]}_{j}
\gamma^{j}\partial^{l}
\Psi^{[n]}_{l}
\\
&{}+(s-2n)(n-1)\,
{ \bar{\Psi}^{[n]}_{l}
\partial^{l}
\slashed{\Psi}^{[n]}}
+\frac{1}{2}(s-1-2n)(s-2n)\,
\bar{\Psi}^{[n]}_{ij}
\gamma^{j}\partial^{i}
\Psi^{[n+1]}
\Bigg\} \, . \label{H_s}
\end{split}
\end{equation}
%

Note that the terms entering the symplectic form and the first class constraints \eqref{F_s_app} contain either zero or two spatial gamma matrices, while the terms entering the Hamiltonian \eqref{H_s} contain either one or three spatial gamma matrices. These properties are relevant to express the minimal magnetic equations of motion in terms of the objects introduced in this appendix.


\bibliographystyle{JHEP}
\bibliography{ref}

\end{document}